\documentclass[12pt]{article}

\ifx\pdfoutput\undefined
\usepackage[dvips,bookmarks]{hyperref}
\else
\usepackage{hyperref}
\fi
\hypersetup{colorlinks=false,bookmarksopen,bookmarksnumbered,citecolor=blue,
   pdfstartview=FitH}

\usepackage{latexsym}
\usepackage{amssymb,amsfonts,amsmath}
\usepackage{graphicx} 
\usepackage{indentfirst}
\usepackage{bbm}
\usepackage{amssymb}
\usepackage{verbatim}
\usepackage{amsmath, amsthm,amssymb}
\usepackage{mathrsfs}
\usepackage{hyperref}
\usepackage{amsfonts}
\usepackage{dsfont}
\usepackage{slashed, tensor}
\usepackage{booktabs}
\usepackage{graphicx}
\usepackage{mathrsfs}

\numberwithin{equation}{section}
\newcommand{\p}[1]{(\ref{#1})}
\newcommand{\be}{\begin{equation}}
\newcommand{\bea}{\begin{eqnarray}}
\newcommand{\ee}{\end{equation}}
\newcommand{\eea}{\end{eqnarray}}

\parskip=\medskipamount

\newcommand{\ga}{\alpha}
\newcommand{\gb}{\beta}
\newcommand{\gc}{\gamma}

\newcommand{\pl}{\partial}

\newcommand{\non}{\nonumber}

\usepackage{hyperref}

\newcommand {\cF}{{\cal F}}

\newcommand {\cJ}{{\cal J}}

\newcommand {\cL}{{\cal L}}
\newcommand {\cM}{{\cal M}}
\newcommand {\cN}{{\cal N}}

\newcommand {\cV}{{\cal V}}
\newcommand {\cW}{{\cal W}}
\newcommand {\cX}{{\cal X}}
\newcommand {\cY}{{\cal Y}}

\def\a{\alpha}
\def\b{\beta}

\def\d{\delta}
\def\e{\epsilon}
\def\f{\phi}
\def\g{\gamma}
\def\G{\Gamma}

\def\k{\kappa}
\def\l{\lambda}

\def\q{\theta}
\def\r{\rho}
\def\s{\sigma}

\def\x{\xi}
\def\z{\zeta}
\def\D{\Delta}
\def\F{\Phi}
\def\J{\Psi}
\def\L{\Lambda}
\def\O{\Omega}

\def\U{\Upsilon}

\def\rd{{\rm d}}
\def\ri{{\rm i}}

\newcommand{\sSL}{\mathsf{SL}}

\newcommand{\1}{{\underline{1}}}
\newcommand{\2}{{\underline{2}}}

\newcommand{\ve}{\varepsilon}
\newcommand{\pa}{\partial}
\newcommand{\hf}{\frac12}

\newcommand{\bm}[1]{\mbox{\boldmath$#1$}}

\begin{document}
\topmargin -1cm \oddsidemargin=0.25cm\evensidemargin=0.25cm
\textwidth 18cm
\setcounter{page}0
\renewcommand{\thefootnote}{\fnsymbol{footnote}}

\begin{titlepage}
\begin{flushright}
September, 2016\\
\end{flushright}
\vskip .4in

\begin{center}
{\Large \bf Off-shell massive \mbox{$\bm{\cN=1}$}  supermultiplets\\ in three dimensions}
\vskip .3in 
{\bf Sergei M. Kuzenko and Mirian Tsulaia}\\
\vskip .2in \footnotesize{
{\it School of Physics M013,
The University of
Western Australia\\
35 Stirling Highway,
Crawley, W.A. 6009, Australia}}\\
\vspace{2mm}
\texttt{sergei.kuzenko@uwa.edu.au, mirian.tsulaia@uwa.edu.au}
\vskip .8in

\begin{abstract}
This paper is mainly concerned with the construction of new off-shell higher spin $\cN=1$ supermultiplets in three spacetime dimensions.
We elaborate on the gauge prepotentials and linearised super-Cotton tensors 
for higher spin $\cN=1$ superconformal geometry
and  propose compensating superfields required to formulate off-shell massless higher spin supermultiplets. The corresponding gauge-invariant actions are worked out explicitly
using an auxiliary oscillator realisation. 
We construct, for the first time, off-shell massive higher spin supermultiplets. 
The gauge-invariant actions for these supermultiplets
are obtained by adding  Chern-Simons like mass terms (that is, higher spin extensions
of the linearised action for $\cN=1$ conformal supergravity) 
to the actions for the massless supermultiplets.
For each of the massive gravitino and supergravity multiplets, 
we propose two dually equivalent formulations.
\end{abstract}

\end{center}

\vfill

\end{titlepage}

\tableofcontents

\renewcommand{\thefootnote}{\arabic{footnote}}

\section{Introduction}

In three spacetime dimensions (3D), the off-shell structure of $\cN=1$ 
supergravity was understood in the late 1970s \cite{HT,BG,Siegel}  
and further elaborated in \cite{GGRS}. 
Since then,  there have appeared a number of important  developments in minimal 
3D supergravity, including the $\cN=1$ topologically massive supergravity 
with and without a cosmological term \cite{DK,Deser}, 
various approaches to $\cN=1$ conformal supergravity 
\cite{vanN85,Uematsu,HIPT,KLT-M11,KT-M12,BKNT-M}, 
3D compactifications of $\cM$-theory with minimal local supersymmetry
(see \cite{Becker} and references therein),
and higher-derivative models 
for massive $\cN=1$ supergravity \cite{ABdeRST,BHRST10,BO,KNT-M}. 
The latter locally supersymmetric theories, which generalise 
the models for massive gravity proposed in \cite{BHT1,BHT2}, 
possess remarkable properties such as unitarity in the presence of
curvature squared terms. Since these massive  theories 
are nonlinear in the  curvature tensor, 
their explicit construction would be extremely difficult to achieve
without making use of the off-shell multiplet calculus for $\cN=1$ supergravity.

The general massive gravity models of  \cite{BHT1,BHT2} and their 
supersymmetric cousins, 
including those proposed in  \cite{ABdeRST,BHRST10,BO,KNT-M}, 
may possess higher spin generalisations, see e.g. \cite{BOT}. 
Surprisingly, to the best of our knowledge,
off-shell massive higher spin $\cN=1$ supermultiplets 
in three dimensions have never been constructed. 
The on-shell massive higher spin 3D $\cN=1$ supermultiplets have been 
formulated recently, both for the Minkowski  and anti-de Sitter (AdS) backgrounds
\cite{BSZ,BSZ2}, building on the elegant gauge-invariant construction of
massive higher spin fields in AdS \cite{Zinoviev}.  
However, since the massive higher spin supermultiplets of \cite{BSZ,BSZ2}
lack auxiliary fields, it could be difficult to use this approach
to generate consistent cubic and possible higher-order couplings 
(as it often happens in supersymmetric field theory). 
The aim of this paper is to construct, for the first time, 
off-shell massive higher spin $\cN=1$ supermultiplets. 

Our paper is a continuation of the recent work \cite{KO} in which 
the off-shell massive higher spin $\cN=2$ supermultiplets were 
constructed in three dimensions. The  structure of these 3D $\cN=2$ 
massive supermultiplets is similar to that of the off-shell 4D $\cN=1$ 
massless supermultiplets \cite{KSP,KS93} (see \cite{BK} for a review)
in the sense that there are two dually equivalent series of off-shell 
formulations. As will be shown below, the 3D $\cN=1$ case 
is more similar to the non-supersymmetric Fronsdal actions \cite{Fronsdal,FF},
for there is essentially a single off-shell formulation for each massive 
higher spin supermultiplet (modulo auxiliary superfields). 
A remarkable feature of our massive $\cN=1$ supermultiplets 
is that they are formulated in terms of {\it unconstrained} superfields, 
unlike their $\cN=2$ counterparts \cite{KO}.
This makes the off-shell higher spin $\cN=1$ supersymmetric theories more tractable 
than the $\cN=2$ ones.  

This paper is organised as follows. In section 2 we define on-shell massive 
superfields and present a manifestly supersymmetric expression for the superhelicity operator. In section 3 we elaborate on  the higher spin superconformal gauge multiplets
and the corresponding gauge invariant field strengths. Section 4 describes 
the massless higher spin gauge prepotentials. The off-shell realisations 
for massless low spin supermultiplets are given in section 5. In section 6 
we present the off-shell massless higher spin supermultiplets, and the massive 
case is presented in section 7. Concluding comments and open problems are discussed in section 8.  The main body of the paper is accompanied by two 
technical appendices.  Our 3D notation and conventions correspond to those introduced in 
\cite{KLT-M11,KPT-MvU}.


\section{Massive (super)fields}

In this section we discuss on-shell (super)fields which realise
the massive representations of the 3D Poincar\'e and 
$\cN=1$ super-Poincar\'e groups. The material in subsection 
\ref{subsection2.1} is taken  almost verbatim from \cite{KO}.

\subsection{Massive fields}\label{subsection2.1}

Let $P_a$ and $J_{ab}= -J_{ba}$ be the generators of the 3D Poincar\'e group.
The Pauli-Lubanski scalar 
\bea
W:= \hf \ve^{abc}P_a J_{bc} = -\hf P^{\a\b} J_{\a\b}
\label{PauliL}
\eea
commutes with the generators $P_a$ and $J_{ab}$.
Irreducible unitary representations of the Poincar\'e group 
are labelled by two parameters,  mass $m>0$ and helicity $\l$, 
which are associated with the Casimir operators, 
\bea
 P^a P_a = -m^2 {\mathbbm 1} ~, \qquad W=m \l {\mathbbm 1}~.
 \label{Casimirs}
 \eea
 One defines $|\l|$ to be the spin. 
 
In the case of field representations, it holds that 
\bea
W= \hf \pa^{\a \b} M_{\a\b}~,
\eea
where the action of the Lorentz generator with spinor indices,  
$M_{\a\b}=M_{\b\a}$,  on a field 
$\f_{\g_1 \cdots \g_n} = \f_{(\g_1 \cdots \g_n)}$ is defined by 
\bea
M_{\a\b} \f_{\g_1 \cdots \g_n} = \sum_{i=1}^n
\ve_{\g_i (\a} \f_{\b) \g_1 \cdots \widehat{\g_i} \dots\g_n}~,
\eea
where the hatted index of $\f_{\b \g_1 \cdots \widehat{\g_i} \dots\g_n}$  is omitted.

For $n>1$, a massive field, $\f_{\a_1 \cdots \a_n} 
= \bar \f_{\a_1\dots \a_n} = \f_{(\a_1 \cdots \a_n)}  $,
is a real symmetric rank-$n$ spinor field
which obeys the differential conditions \cite{TV} (see also \cite{BHT})
\begin{subequations}\label{2.55}
\bea
\pa^{\b\g} \f_{\b\g\a_1 \cdots \a_{n-2}} &=&0~, \label{dif_sub} \\
\pa^\b{}_{(\a_1} \f_{\a_2 \dots \a_n)\b} &=& m \s \f_{\a_1\dots \a_n}~,
\qquad \s =\pm 1~.
\label{mass3.6}
\eea 
\end{subequations}
In the spinor case, $n=1$, eq.  \eqref{dif_sub} is absent, and 
 the massive field is defined to obey the Dirac equation 
\eqref{mass3.6}.
It is easy to see that  \eqref{dif_sub} and  \eqref{mass3.6}
imply the mass-shell equation\footnote{The equations \eqref{dif_sub} and 
\eqref{mass-shell} prove to be equivalent to the 3D Fierz-Pauli field equations \cite{FP}.}
\bea
(\Box -m^2 ) \f_{\a_1 \cdots \a_n} =0~,
\label{mass-shell}
\eea
which is the first equation in \eqref{Casimirs}. In the spinor case, $n=1$,
eq. \eqref{mass-shell} follows from the Dirac equation \eqref{mass3.6}. 
The second relation in \eqref{Casimirs} also holds, with 
\bea
\l = \frac{n}{2} \s~.
\eea 
The spin of $\f_{\a(n)} $ is $n/2$.


\subsection{Massive $\cN=1$ superfields}\label{section2.2}

Let $P_{a}$,  $J_{ab}= -J_{ba}$, $Q_\a$ 
 be the generators of the 3D $\cN=1$ super-Poincar\'e group.
 The supersymmetric extension of the Pauli-Lubanski scalar 
 \eqref{PauliL} is the following operator  \cite{MT}
\bea
Z= W-\frac{\ri}{8} Q^2
= \hf \ve^{abc}P_a J_{bc} -\frac{\ri}{8} Q^\a  Q_\a ~,
\label{super-helicity}
\eea
which commutes with the supercharges,
\bea
[Z,Q_\a ] =0~.
\eea
The operator $Z$ is analogous to the 4D $\cN=1$  superhelicity operator
 introduced in  \cite{BK}.
Irreducible unitary representations of the $\cN=1$ super-Poincar\'e group 
are labelled by two parameters,  mass $m$ and superhelicity $\k$, 
which are associated with the Casimir operators, 
\bea
 P^a P_a = -m^2 {\mathbbm 1} ~, \qquad Z=m  \k {\mathbbm 1}~.
 \label{Casimirs-super}
 \eea
 Our definition of 
 the superhelicity 
 agrees with 
 \cite{MT}.
 The massive representation of superhelicity $\k$ is a direct sum of 
 two massive  representations
 of the Poincar\'e group with helicity values $(\k-\frac{1}{4}, \k+ \frac{1}{4})$. 
If $\k -\frac{1}{4} $ is not an integer, the supermultiplet describes anyons.
The case $2 \k \in {\mathbb Z}$ corresponds to the so-called semion supermultiplets, 
of which  the $\k=\hf $ supermultiplet was first studied in \cite{SV}.

When dealing with the supermultiplets containing particles of (half-)integer helicity, it appears more convenient, by analogy with the 
 $\cN=1$ case in four dimensions \cite{BK}, to define a shifted 
superhelicity operator, $\hat \k = \k- \frac{1}{4}$, which takes integer or half-integer values.  
However, here we will use the definition introduced in \cite{MT}.

In the case of superfield representations of  the $\cN=1$ super-Poincar\'e group, 
the  infinitesimal super-Poincar\'e transformation of a tensor superfield is 
\bea
 \d \F =\ri (-b^a P_a +\hf \L^{ab} J_{ab} + \e^\a Q_\a) \F
 =\ri \left(\hf b^{\a\b} P_{\a\b}  +\hf \L^{\a \b} J_{\a \b} + \e^\a Q_\a \right)\F~,
 \eea
 where the generators of spacetime translations ($P_{\a\b}$), 
 Lorentz transformations ($\L_{\a\b}$) and supersymmetry transformations 
 ($Q_\a$) are 
 \begin{subequations}
 \bea
 P_{\a\b}&=& -\ri \pa_{\a\b} ~,\qquad \pa_{\a\b}= (\g^m)_{\a\b} \pa_m~, \\
 J_{\a\b} &=& \ri \q_{(\a} \pa_{\b)} -\ri M_{\a\b}~,\\
 Q_\a &=& \pa_\a +\ri \q^\b \pa_{\a\b} ~,\qquad \pa_\a =\frac{\pa}{\pa \q^\a}~.
 \eea
 \end{subequations} Using the explicit expressions for the 
 super-Poincar\'e generators, 
 the superhelicity operator \eqref{super-helicity} can be written in a
 manifestly supersymmetric form 
 \bea
 Z =\hf \pa^{\a \b} M_{\a\b} -\frac{\ri}{8} D^2~.
 \eea

For $n>0$, a massive superfield $T_{\a(n)}$
is defined to be a real symmetric rank-$n$ spinor,
 $T_{\a_1 \cdots \a_n} 
= \bar T_{\a_1\dots \a_n} = T_{(\a_1 \cdots \a_n)}  $,
which obeys the differential conditions \cite{KNT-M}
\begin{subequations}
\label{214}
\bea
D^\b T_{\b \a_1 \cdots \a_{n-1}} &=& 0 \quad \Longrightarrow \quad
\pa^{\b\g} T_{\b\g\a_1\dots \a_{n-2}} =0
~ , 
\label{214a} 
\\
-\frac{\ri}{2} D^2  T_{\a_1 \dots \a_n} &=& m \s T_{\a_1 \dots \a_n}~, 
\qquad \s =\pm 1~.
\label{214b}
\eea
\end{subequations}
It follows from \eqref{214a} that 
\bea
-\frac{\ri}{2} D^2  T_{\a_1 \dots \a_n} =\pa^\b{}_{(\a_1} T_{\a_2 \dots \a_n)\b} ~,
\eea
and thus $T_{\a(n)}$ is an on-shell superfield, 
\bea
\pa^\b{}_{(\a_1} T_{\a_2 \dots \a_n)\b} = m \s T_{\a_1 \dots \a_n}~,
\qquad \s =\pm 1~.
\eea
Making use of  the identity \eqref{iden4},
we also deduce directly from \eqref{214b} that\footnote{The equations 
\eqref{214a} and \eqref{2177} provide the $\cN=1$ supersymmetric extensions 
of the 3D Fierz-Pauli equations.} 
\bea
(\Box -m^2) T_{\a(n)}=0~.
\label{2177}
\eea
For the superhelicity of $T_{\a(n)}$  we obtain 
\bea
\k = \hf\left( n +\hf \right) \s~.
\label{218}
\eea
We define the superspin of $T_{\a(n)}$ to be $n/2$. 
The massive supermultiplet $T_{\a(n)}$ contains two ordinary 
massive fields of the type \eqref{2.55}, which are
\bea
\f_{\a_1 \dots \a_n} := T_{\a_1 \dots \a_n} |_{\q=0}~, \qquad 
\f_{\a_1 \dots \a_{n+1}} := \ri^{n+1} D_{(\a_1} T_{\a_2 \dots \a_{n+1})} |_{\q=0}~.
\eea
Their helicity values are $\frac{n}{2} \s $ and $\frac{n+1}{2}  \s$, respectively.

As an example, let us consider the following model for a massive scalar multiplet 
\bea
S_{\rm SM} [X]=
-\frac{\ri}{2} \int \rd^{3|2} z \,  D^\a X D_\a X
+m \s \int \rd^{3|2} z \,  X^2~, \qquad \s =\pm 1~.
\label{222}
\eea
Throughout this paper, the $\cN=1$ superspace integration measure\footnote{This 
definition implies that $ \int \rd^{3|2} z \, V = \int \rd^3 x\, F$, for any scalar 
superfield $V(x,\q)  =\dots + \ri \q^2 F(x)$.}
is defined as follows:
\bea
\int \rd^{3|2} z \, L= \frac{\ri}{4} \int \rd^3 x \, D^2 L\Big|_{\q=0}~.
\eea
The equation of motion for the action \eqref{222} is 
\bea
-\frac{\ri}{2} D^2 X = m \s X ~,
\eea
which shows that the superhelicity of $X$ is $\k = \frac{1}{4} \s$.


\section{$\cN=2 \to \cN=1$ superspace reduction: 
Superconformal  gauge multiplets}

In general, off-shell $\cN=1$ higher spins supermultiplets in three dimensions
may be obtained 
by applying the $\cN=2 \to \cN=1$ superspace reduction to the 
$\cN=2$ supermultiplets constructed in \cite{KO}. We denote by 
${\mathbb D}_\a$ and $\bar {\mathbb D}_\a$ the spinor covariant derivatives
of the $\cN=2$ Minkowski superspace ${\mathbb M}^{3|4}$. They obey the 
anti-commutation relations
\bea
\{{\mathbb D}_\a, \bar {\mathbb D}_\b\}=-2\ri\, \pa_{\a\b}~,\qquad
\{{\mathbb D}_\a,{\mathbb D}_\b\}=\{ \bar {\mathbb D}_\a, \bar {\mathbb D}_\b\}=0~.
\label{N=2acd}
\eea
In order to carry out the $\cN=2 \to \cN=1$ superspace reduction, 
it is useful to introduce 
real Grassmann coordinates  $\q^\a_I$ for ${\mathbb M}^{3|4}$, 
where $I =\1, \2$. We define these coordinates 
by choosing the corresponding spinor covariant derivatives
$D^I_\a$ as in \cite{KPT-MvU}:
\bea
&&
{\mathbb D}_\a=\frac{1}{ \sqrt{2}}(D_\a^{\1}-\ri D_\a^{\2})~,\qquad
\bar {\mathbb D}_\a=-\frac{1}{ \sqrt{2}}(D_\a^{\1}+\ri D_\a^{\2})~.~~~
\label{2.1}
\eea
From \eqref{N=2acd} we deduce
\bea
\big\{ D^I_\a , D^J_\b \big\} = 2{\rm i}\, \d^{IJ}  (\g^m)_{\a\b}\,\pa_m~, 
\qquad I ,J=\1, \2~.
\eea

Given an $\cN=2$ superfield $U(x, \q_I)$, we define its $\cN=1$ bar-projection
\bea
U|:= U(x, \q_I)|_{\q_{\2} =0}~.
\eea
It is clear that $U|$ is a superfield on $\cN=1$ Minkowski superspace  
${\mathbb M}^{3|2}$ parametrised
by real Cartesian coordinates $z^A= (x^a, \q^\a)$, where $\q^\a:=\q^\a_{\1}$.
The covariant derivative of $\cN=1$ Minkowski superspace
$D_\a := D_\a^{\1}$
obeys the anti-commutation relation
\bea
\big\{ D_\a , D_\b \big\} = 2{\rm i}\,   (\g^m)_{\a\b}\,\pa_m~.
\eea


\subsection{Higher spin superconformal gauge multiplets}\label{subsection3.1}

In accordance with \cite{KO},
the higher spin $\cN=2$ superconformal  gauge multiplet 
is described in terms 
 of a real unconstrained prepotential 
 \bea
 {\mathbb H}_{\a(n)} :={\mathbb H}_{\a_1\dots \a_n} ={\mathbb H}_{(\a_1\dots \a_n)} 
 = \bar  {\mathbb H}_{\a(n)}~, 
  \eea
which is defined modulo gauge transformations of the form 
\begin{subequations} \label{2.7}
\bea
\d {\mathbb H}_{\a(n)} = g_{\a(n)} + \bar g_{\a(n)}~, 
\label{2.7a}
\eea
where the complex gauge parameter  
$g_{\a(n)} =g_{\a_1\dots \a_n} =g_{(\a_1\dots \a_n)} $ 
is a longitudinal linear superfield constrained by
\bea
\bar{\mathbb D}_{(\a_{1}}g_{\a_{2}...\a_{n+1})}  =0 
\qquad \Longrightarrow \qquad \bar{\mathbb D}^2
g_{\a (n)}  =0
~.
\label{longlin}
\eea
\end{subequations}
This constraint can always be solved in terms of a complex 
unconstrained potential  $L_{\a(n-1)}$ by the rule
\bea
g_{\a_1\dots \a_n} = \bar{\mathbb D}_{(\a_{1}}L_{\a_{2}...\a_{n})} ~.
\eea
However we will not use this representation in the present paper. 
 
Making use of the representation \eqref{2.1}, the longitudinal linear constraint 
\eqref{longlin} takes the form
\bea
D^{\2}{}_{(\a_1}g_{\a_{2}...\a_{n+1})}  = \ri D^{\1}{}_{(\a_{1}}g_{\a_{2}...\a_{n+1})} ~.
\label{2.9}
\eea
This tells us that, upon reduction to $\cN=1$ superspace, 
$g_{\a(n)}$
is equivalent to two complex unconstrained $\cN=1$ superfields, which 
are obtained by Taylor-expanding  the $\cN=2$ superfield 
$g_{\a(n)}(\q_I) =g_{\a(n)}(\q_\1, \q_\2)$ in powers 
of $\q^\a_{\2}$
and which 
may be chosen as
\bea
g_{\a_1 \dots \a_n}|~, \qquad D^{\2 \,\b} g_{\a_1 \dots \a_{n-1} \b}|~.
\label{2.10}
\eea
Upon reduction to $\cN=1$ superspace, 
the gauge prepotential $ {\mathbb H}_{\a(n)} $ is equivalent 
to four unconstrained superfields
\bea
 {\mathbb H}_{\a_1 \dots \a_n} |~, \quad
D^{\2}{}_{(\a_1} {\mathbb H}_{\a_2 \dots \a_{n+1})}| ~,\quad
 D^{\2 \,\b} {\mathbb H}_{\a_1 \dots \a_{n-1}\b}|~, \quad 
 \frac{\ri}{4} (D^{\2})^2{\mathbb H}_{\a_1 \dots \a_n}|~.
\eea
Here the first and the fourth superfields are real, 
while the other superfields are real or imaginary depending on $n$.

Since the $\cN=1$ gauge parameters \eqref{2.10} are complex unconstrained, 
it is in our power to choose the $\cN=1$ supersymmetric gauge conditions
\bea
{\mathbb H}_{\a_1 \dots \a_n} |=0~,
 \qquad  D^{\2 \,\b} {\mathbb H}_{\a_1 \dots \a_{n-1}\b}|=0~.
 \label{2.12}
\eea
In this gauge we stay with the following real unconstrained
$\cN=1$ superfields 
\begin{subequations} \label{2.13}
\bea
H_{\a_1 \dots \a_{n+1} } &:=& 
\ri^{n+1}
D^{\2}{}_{(\a_1} {\mathbb H}_{\a_2 \dots \a_{n+1})}|~, \label{2.13a} \\
H_{\a_1\dots \a_n} &:=& \frac{\ri}{4} (D^{\2})^2{\mathbb H}_{\a_1 \dots \a_n}|~.
\label{2.13b}
\eea
\end{subequations}
The residual gauge freedom, which preserves the gauge conditions \eqref{2.12},
 is described by real unconstrained $\cN=1$ 
superfield parameters $\z_{\a(n)}$ and $\z_{\a(n-1)}$ defined by 
\begin{subequations} \label{2.14}
\bea
g_{\a_1 \dots \a_n }| &=&-\frac{\ri}{2} \z_{\a_1 \dots \a_n}~, 
\qquad \bar \z_{\a(n)} = \z_{\a(n)}~, 
\label{2.14a} \\
D^{\2 \,\b} g_{\a_1 \dots \a_{n-1}\b}| &=&
-\ri^{n}
\frac{n+1}{n}
 \z_{\a_1 \dots \a_{n-1} }~, 
\qquad \bar \z_{\a(n-1)} = \x_{\a(n-1)}~, \label{2.14b}
\eea
\end{subequations}
This leads to 
\bea
\d H_{\a_1 \dots \a_{n+1} } 
& \propto &
D^{\2}{}_{(\a_1}  \d {\mathbb H}_{\a_2 \dots \a_{n+1})}|
= D^{\2}{}_{(\a_1}  g_{\a_2 \dots \a_{n+1})}|
+D^{\2}{}_{(\a_1}  \bar g_{\a_2 \dots \a_{n+1})}| \nonumber \\
&=& \ri D^{\1}{}_{(\a_1}  g_{\a_2 \dots \a_{n+1})}|
- \ri D^{\1}{}_{(\a_1}  \bar g_{\a_2 \dots \a_{n+1})}|
=  D{}_{(\a_1}  \z_{\a_2 \dots \a_{n+1})}~,~~~~~
\eea
where we have used the longitudinal linear constraint
\eqref{2.9} and the explicit expression  \eqref{2.14a}
for the residual gauge transformation.
The final result for the gauge transformation of \eqref{2.13a} is
\begin{subequations}
\bea
\d H_{\a_1 \dots \a_{n+1} } 
= \ri^{n+1}
D{}_{(\a_1}  \z_{\a_2 \dots \a_{n+1})}~.
\label{2.16}
\eea
In a similar way we determine the gauge transformation of \eqref{2.13b} to be 
\bea
\d H_{\a_1 \dots \a_{n} } 
= \ri^{n}
D{}_{(\a_1}  \z_{\a_2 \dots \a_{n})}~.
\label{2.17}
\eea
\end{subequations}
This agrees with \eqref{2.16} if we replace $n \to n+1$.
The superconformal prepotential $H_{\a(n)}$ and its gauge transformation 
\eqref{2.17} were introduced  in \cite{K16}.

In discussing $\cN=1$ superconformal multiplets, we follow
the formalism described in 
\cite{KPT-MvU,BKS}.
The $\cN=1$ superconformal transformations are generated by  
conformal Killing supervector field.
\be
\x  = \x^{a}   \pa_{ a} 
+  \x^{\a}   D_{\a} ~.
\ee
By definition, the $\cN=1$ conformal Killing supervector field
obeys the equation $[\x, D_\a ] \propto D_\b$, or equivalently
\bea
[\x, D_\a ] = 
-K_\a{}^\b  D_\b -\hf \s  D_\a~,
\label{2.19}
\eea
which implies
\begin{subequations}
\bea
& &\x^\a = \frac{\ri}{6} D_\b \x^{\b\a}~,  \\
&& D_{(\g} \x_{\a\b)} =0~,
\label{scK}
\eea 
\end{subequations}
of which \eqref{scK} 
is the $\cN=1$ superconformal Killing equation.
In \eqref{2.19} we have introduced the $z$-dependent parameters of Lorentz ($K_{\a \b}$) and
scale ($\s$) transformations 
\bea
K_{\a\b} &:=&  D_{(\a} \x_{\b)} 
~, \qquad
\s:= D_\a \x^\a =\frac{1}{3} \pa_a \x^a~.
\eea
These parameters are related to each other by the relation
\bea
D_\a K_{\b \g} &=&-  \ve_{ \a ( \b} D_{ \g)} \s~,  
\eea
which implies 
\be
D^2 \s=0~.
\ee

A symmetric rank-$n$ spinor superfield $\F_{\a(n)}=\F_{\a_1 \dots \a_n} $ 
is said to be primary of dimension $d_\F$ 
if its superconformal transformation is 
\bea
\d_\x \F_{\a_1 \dots \a_n} = \x \F_{\a_1 \dots \a_n} 
+n K^\b{}_{(\a_1} \F_{\a_2 \dots \a_n)\b } + d_{\F} \s \F_{\a_1 \dots \a_n} ~.
\eea 
We now require both the gauge field $H_{\a(n)} $ and the gauge parameter
$\z_{\a(n-1)}$ in \eqref{2.17} to be primary superfields.
This is consistent if and only if the dimension of $H_{\a(n)} $ is equal to 
 $(1-n/2)$, as stated in  \cite{K16}. Thus the superconformal transformation law of 
 $H_{\a(n)}$ is 
\bea
\d_\x H_{\a_1 \dots \a_n} = \x H_{\a_1 \dots \a_n} 
+n K^\b{}_{(\a_1} H_{\a_2 \dots \a_n)\b } + (1-\frac{1}{2} n) \s H_{\a_1 \dots \a_n} ~.
\label{2.24}
\eea 


\subsection{Higher spin superconformal field strengths}

To start with, it is worth recalling the 
$\cN=2$ superconformal gauge-invariant field strength, 
${\mathbb W}_{\a(n)} = \bar {\mathbb W}_{\a(n)}$, introduced in \cite{KO}
\bea
&&{\mathbb W}_{\a_1 \dots \a_n} 
:= \frac{1}{2^{n}} 
\sum\limits_{J=0}^{\left \lfloor{n/2}\right \rfloor}
\bigg\{
\binom{n}{2J} 
\Delta  \Box^{J}\pa_{(\a_{1}}{}^{\b_{1}}
\dots
\pa_{\a_{n-2J}}{}^{\b_{n-2J}}
{\mathbb H}_{\a_{n-2J+1}\dots\a_{n})\b_1 \dots\b_{n-2J}}~~~~
\nonumber \\
&&\qquad \qquad +
\binom{n}{2J+1}\Delta^{2}\Box^{J}\pa_{(\a_{1}}{}^{\b_{1}}
\dots\pa_{\a_{n-2J -1}}{}^{\b_{n-2J -1}}
{\mathbb H}_{\a_{n-2J}\dots\a_{n})
\b_1 \dots \b_{n-2J -1} }\bigg\}~,~~~~~
\label{2.25}
\eea
where $\left \lfloor{x}\right \rfloor$ denotes the floor (or the integer part) of 
a number $x$, and the operator $\D$ is
\bea
 \D =\frac{\ri}{2} {\mathbb D}^\a \bar {\mathbb D}_\a  
 = \frac{\ri}{2} \bar {\mathbb D}^\a  {\mathbb D}_\a  ~.
  \eea
There are three fundamental properties  
that ${\mathbb W}_{\a(n)}$ possesses.
Firstly, it is invariant under the gauge transformations \eqref{2.7}. 
Secondly, it obeys the Bianchi identity
 \bea
{\mathbb D}^{\b} {\mathbb W}_{\b\a_1 \dots \a_{n-1}}=0 
\quad \Longleftrightarrow \quad
\bar{\mathbb D}^{\b}{\mathbb W}_{\b\a_1 \dots \a_{n-1}}=0 ~.
\label{2.27}
\eea
Thirdly, the real symmetric rank-$n$ spinor 
  ${\mathbb W}_{\a(n)}$ is a primary $\cN=2$ superfield of dimension 
 $(1+n/2)$. As explained in \cite{KO}, the conditions that 
  ${\mathbb W}_{\a(n)}$ is primary and obeys the constraints \eqref{2.27} 
  are consistent if and only if the dimension of  ${\mathbb W}_{\a(n)}$ is equal to 
  $(1+n/2)$.
 If the prepotential ${\mathbb H}_{\a(n)}$ is chosen to be 
 primary of dimension $(-n/2)$, then its descendant \eqref{2.25} proves
 to be primary of dimensions $(1+n/2)$. It is important to emphasise 
 that the most general solution to the constraints \eqref{2.27} is given by 
 \eqref{2.25}, as discussed in \cite{K16}.

In the $n=2$ case, the field strength $W_{\a\b}(H)$ coincides with the linearised
version \cite{CDFKS,KO} of the $\cN=2$ super-Cotton tensor 
\cite{K12,BKNT-M}. Thus the field strength \eqref{2.25} for $n>2$
is the higher-spin extension of 
the super-Cotton tensor.

We now turn to reducing the field strength ${\mathbb W}_{\a(n)}$ 
to $\cN=1$ superspace. In the real basis for the $\cN=2$ spinor covariant derivatives,
the Bianchi identities \eqref{2.27} read
\bea
{D}^{I \b} {\mathbb W}_{\b\a_1 \dots \a_{n-1}}=0~, \qquad  I =\1, \2~.
\eea
These constraints imply that, upon reduction to $\cN=1$ superspace, 
${\mathbb W}_{\a(n)}$ 
is equivalent to the following real $\cN=1$ superfields
\bea
{W}_{\a_1 \dots \a_{n}}:= {\mathbb W}_{\a_1 \dots \a_{n}}|~, \qquad
{W}_{\a_1 \dots \a_{n+1}} \propto 
\ri^{n+1} D^{\2}{}_{(\a_1} {\mathbb W}_{\a_2 \dots \a_{n+1})}|~,
\eea
each of which is divergenceless, in particular 
\bea
{D}^{\b} {W}_{\b\a_1 \dots \a_{n-1}}=0 ~.
\label{2.30}
\eea
We now compute the bar-projection of \eqref{2.25} in the gauge \eqref{2.12}
and make use of the identities
\bea
\D = -\frac{\ri}{4} \Big\{ (D^{\1})^2 + (D^{\2})^2 \Big\} ~,\qquad
\D^2  =\frac{1}{8}\Big\{4 \Box -  (D^{\1})^2  (D^{\2})^2 \Big\}~.
\eea
Making use of these identities  leads to 
the $\cN=1$  field strength\footnote{It was given without derivation in \cite{K16}.}
\bea
&&W_{\a_1 \dots \a_n} (H)
:= \frac{1}{2^{n}} 
\sum\limits_{J=0}^{\left \lfloor{n/2}\right \rfloor}
\bigg\{
\binom{n}{2J}  \Box^{J}\pa_{(\a_{1}}{}^{\b_{1}}
\dots
\pa_{\a_{n-2J}}{}^{\b_{n-2J}}H_{\a_{n-2J+1}\dots\a_{n})\b_1 \dots\b_{n-2J}}~~~~
\nonumber \\
&&\qquad \qquad -\frac{\ri}{2} 
\binom{n}{2J+1}D^{2}\Box^{J}\pa_{(\a_{1}}{}^{\b_{1}}
\dots\pa_{\a_{n-2J -1}}{}^{\b_{n-2J -1}}H_{\a_{n-2J}\dots\a_{n})
\b_1 \dots \b_{n-2J -1} }\bigg\}~.~~~~~
\label{2.32}
\eea
This real superfield, $W_{\a(n)}= \bar W_{\a(n)}$, is a descendant of 
 the real unconstrained prepotential $H_{\a(n)}$ defined 
 modulo the gauge transformations \eqref{2.17}.
 The field strength proves to be  gauge invariant,  
 \bea
\d H_{\a_1 \dots \a_{n} } = \ri^{n} D{}_{(\a_1}  \z_{\a_2 \dots \a_{n})}
\quad \Longrightarrow \quad 
  \d W_{\a(n)}=0~, 
\eea
and  obey the Bianchi identity \eqref{2.30}.
Using the superconformal transformation law of $H_{\a(n)}$, eq. \eqref{2.24},
one may check that  the superconformal transformation law of
the field strength \eqref{2.32} is
\bea
\d_\x W_{\a_1 \dots \a_n} = \x W_{\a_1 \dots \a_n} 
+n K^\b{}_{(\a_1} W_{\a_2 \dots \a_n)\b } + (1+\frac{1}{2} n) \s W_{\a_1 \dots \a_n} ~,
\label{2.34}
\eea 
and therefore $W_{\a(n)}$ is a primary superfield of dimension $(1+n/2)$.

For $n=1$ the field strength \eqref{2.32} is 
\bea
2W_\a  = -\pa_\a{}^\b H_\b +\frac{\ri}{2} D^2 H_\a = \ri D^\b D_\a H_\b~,
\label{3.36}
\eea
as a consequence of the anti-commutation relation \eqref{ALG}
The final expression for $W_\a$ in \eqref{3.36}
coincides with the gauge-invariant field strength 
of a vector multiplet \cite{GGRS}. 
For $n=2$ the field strength $W_{\a\b}$ given by \eqref{2.32} can be seen to coincide with the gravitino field 
strength \cite{GGRS}. Finally, for  $n=3$ the field strength $W_{\a\b\g}$ 
given by \eqref{2.32} is
the linearised version  \cite{KNT-M} of the $\cN=1$ super-Cotton tensor 
\cite{KT-M12,BKNT-M}. At the component level, field strength \eqref{2.32} 
for $n=2s$ contains (as the $\q$-independent component)
the bosonic higher spin Cotton tensors proposed by Pope
and Townsend \cite{PopeTownsend}, as shown in \cite{K16}. In the 
$n=2s+1$ case, the fermionic ($\q$-independent) component of $W_{\a(2s+1})$ 
was given in \cite{K16}.
The fermionic component of $W_{\a(3)}$, known as the 
Cottino tensor, was first introduced in \cite{ABdeRST}.

It should be pointed out that \eqref{2.32} is the most general solution of the constraint
\eqref{2.30}, as was emphasised in \cite{K16}.
The simplest way to prove this is the observation that 
the field strength \eqref{2.32} may be recast in the form\footnote{The numerical 
coefficient in the right-hand side of \eqref{3366} was not computed in \cite{K16}.} 
\cite{K16}
\bea
W_{\a(n)} = \frac{(-\ri)^n}{2^n} 
D^{\b_1} D_{\a_1} \dots D^{\b_n} D_{\a_n} H_{\b_1 \dots \b_n}~.
\label{3366}
\eea
It is completely symmetric, $W_{\a_1 \dots \a_n} =W_{(\a_1 \dots \a_n)}$,
as a consequence of \eqref{DDDD}. If the superconformal prepotential is constrained 
to be transverse, $D^\b H_{\b \a(n-1)} =0$, the expression for the super-Cotton 
 simplifies, 
\begin{subequations} \label{3377}
\bea
 D^\b H_{\b \a_1 \dots \a_{n-1}}=0 \quad \Longrightarrow \quad 
 W_{\a(n)} = \pa_{\a_1}{}^{\b_1} \dots \pa_{\a_n}{}^{\b^n} H_{\b_1 \dots \b_n}~.
\eea
This result can be fine-tuned as follows:
\bea
W_{\a(2s)} &=&\Box^s H_{\a(2s)}~, \\
W_{\a(2s+1)} &=&\Box^s \pa^\b{}_{(\a_1} H_{\a_2 \dots \a_{2s+1})\b }
=\Box^s \pa^\b{}_{\a_1 } H_{\a_2 \dots \a_{2s+1} \b }~.
\eea
\end{subequations}

Associated with $W_{\a(n)} (H)$ is the gauge-invariant Chern-Simons action 
\bea
S_{\rm CS}[H] = \ri^n \int \rd^{3|2}z \,H^{\a(n)} W_{\a(n)}(H)~,
\label{3.37}
\eea
which is also invariant under the superconformal transformations  \eqref{2.24}.
The action \eqref{3.37} coincides for $n=1$ with the topological mass term for the Abelian vector multiplet  \cite{Siegel}. In the $n=3$ case, \eqref{3.37}
proves to be the linearised action for $\cN=1$ conformal supergravity, 
as may be shown using the results in \cite{GGRS,KT-M12}. 


\section{$\cN=2 \to \cN=1$ superspace reduction: 
Massless gauge multiplets}

There are two series of the massless half-integer superspin 
$\cN=2$ multiplets \cite{KO}, which are dual to each other. 
Here we describe  their $\cN=2 \to \cN=1$ superspace reduction.
Throughout this section, we fix an integer $s>1$. 

\subsection{Longitudinal formulation}

The longitudinal formulation is realised in terms of the following  dynamical variables:
\begin{align}
\mathcal{V}^{\parallel}=&\big\{ {\mathbb H}_{\a(2s)},\,{\mathbb G}_{\a(2s-2)},
\,\bar{\mathbb G}_{\a(2s-2)} \big\} ~,
\label{TL4.1}
\end{align}
where the real superfield ${\mathbb H}_{\a(2s)} ={\mathbb H}_{(\a_{1}\dots \a_{2s})}$ is unconstrained, 
and the complex superfield ${\mathbb G}_{\a(2s-2)}= {\mathbb G}_{(\a_1 \dots \a_{2s-2})}$ is  longitudinal linear,  
\bea
\bar{\mathbb D}_{(\a_{1}}{\mathbb G}_{\a_{2}\dots\a_{2n-1})}  =0~.
\label{4.1}
\eea
The dynamical superfields are defined modulo 
gauge transformations of the form 
\begin{subequations} \label{4.3}
\bea
\delta {\mathbb H}_{\a_1 \dots \a_{2s}}&=& g_{\a_1 \dots \a_{2s}}
+\bar{g}_{\a_1 \dots \a_{2s}}~,  \label{4.3a}\\
\delta {\mathbb G}_{\a_1 \dots \a_{2s-2}}&=&
\frac{s}{2s+1}{\mathbb D}^{\b}\bar{\mathbb D}^{\g}g_{\b \g \a_1 \dots \a_{2s-2}}
+\ri s\pa^{\b \g}g_{\b \g \a_1 \dots \a_{2s-2}}~, \label{4.3b}
\eea
\end{subequations}
where the complex gauge parameter  
$g_{\a_1 \dots \a_{2s}} =g_{(\a_1 \dots \a_{2s})} $ 
is an arbitrary longitudinal linear superfield, eq. \eqref{longlin}.
Clearly, $ {\mathbb H}_{\a(2s)}$ is the higher spin superconformal gauge multiplet 
with $n=2s$ introduced in section \ref{subsection3.1}.
The superfields ${\mathbb G}_{\a(2s-2)}$ and 
$\bar{\mathbb G}_{\a(2s-2)} $ should be viewed as  compensators.
The gauge-invariant action is 
\begin{align}
S^{\parallel}_{s+\frac{1}{2}}[{\mathbb H},{\mathbb G}, \bar {\mathbb G}]
=& \Big(-\frac{1}{2}\Big)^{s} \int \rd^{3|4}z\,
\bigg\{\frac{1}{8}{\mathbb H}^{\a(2s)}D^{\gamma}\bar{D}^{2}D_{\gamma}
{\mathbb H}_{\a(2s)}
\nonumber &\\
&-\frac{1}{16}\Big( [D_{\b_{1}},\bar{D}_{\b_{2}}]{\mathbb H}^{\b_1 \b_2\a(2s-2)}\Big)
[D^{\gamma_{1}},\bar{D}^{\gamma_{2}}]{\mathbb H}_{\gamma_1 \g_2\a(2s-2)}
\nonumber &\\
&+\frac{s}{2} \Big(\pa_{\b_{1}\b_{2}}{\mathbb H}^{\b_1\b_2\a(2s-2)}\Big)
\pa^{\gamma_{1}\gamma_{2}}{\mathbb H}_{\gamma_1 \g_2\a(2s-2)}\nonumber &\\
&+\ri\frac{2s-1}{2s}\left({\mathbb G}-\bar{\mathbb G}\right)^{\a(2s-2)}
\pa^{\b_1\b_2}{\mathbb H}_{\b_1\b_2\a(2s-2)}\nonumber &\\
&+\frac{2s-1}{2s^2}{\mathbb G}\cdot \bar{\mathbb G}
-\frac{2s+1}{4s^2}\left({\mathbb G}\cdot {\mathbb G}
+\bar{\mathbb G} \cdot \bar {\mathbb G}\right)\bigg\}~.
\label{N2LongAction}
&
\end{align}

The $\cN=2 \to \cN=1$ superspace reduction of the superconformal gauge multiplet 
$ {\mathbb H}_{\a(2s)}$ was carried out in section \ref{subsection3.1}. 
It remains to reduce  the compensator 
${\mathbb G}_{\a(2s-2)}$ to $\cN=1$ superspace. 
From the point of view of $\cN=1$ supersymmetry, 
${\mathbb G}_{\a(2s-2)}$ is equivalent to two {\it complex unconstrained} superfields,
which we define as follows:
\bea
{G}_{\a_1 \dots \a_{2s-2}}:={\mathbb G}_{\a_1 \dots \a_{2s-2}}|~, \qquad
\O_{\a_1 \dots \a_{2s-3}}:= \ri  D^{\2\, \b} {\mathbb G}_{\b \a_1 \dots \a_{2s-3}}|~.
\label{4.4}
\eea
Making use of the gauge transformation \eqref{4.3b} gives 
\begin{subequations} \label{4.5}
\bea 
\delta {\mathbb G}_{\a_1 \dots \a_{2s-2}}&=&
\frac{2 \ri s^2}{2s+1} \pa^{\b\g} g_{\b \g \a_1 \dots \a_{2s-2}}
-\frac{\ri s}{2s+1} D^{\1 \, \b} D^{\2\, \g} g_{\b \g \a_1 \dots \a_{2s-2}}~, \\
 D^{\2\, \b} \d{\mathbb G}_{\b \a_1 \dots \a_{2s-3}}&=&
\frac{2 \ri s^2}{2s+1} \pa^{\b\g} D^{\2 \, \d} g_{\b \g \d \a_1 \dots \a_{2s-3}}
-\frac{ s}{2s+1} D^{\1 \, \b} \pa^{\g \d} g_{\b \g \d \a_1 \dots \a_{2s-3}}~.~~~
\eea
\end{subequations}
At this stage one should recall that upon imposing the 
$\cN=1$ supersymmetric gauge conditions \eqref{2.12} the residual gauge freedom
is described by the gauge parameters \eqref{2.14a} and \eqref{2.14b}.
From \eqref{4.5} we read off the gauge transformations of the $\cN=1$ 
complex superfields \eqref{4.4}
\begin{subequations}\label{4.6}
\bea
\d G_{\a(2s-2)} &=& \frac{s^2}{2s+1} \pa^{\b\g} \z_{\b\g \a(2s-2)} 
-(-1)^s \frac{\ri}{2} D^\b \z_{\b \a(2s-2)}~, \\
\d \O_{\a(2s-3)} &=& - \frac{s}{2(2s+1)} D^\b \pa^{\g\d} \z_{\b\g\d \a(2s-3)}
+(-1)^s s \pa^{\b \g} \z_{\b\g \a(2s-3)}~.
\eea
\end{subequations}

In the $\cN=1$ supersymmetric gauge \eqref{2.12}, $ {\mathbb H}_{\a(2s)}$
is described by the two real unconstrained superfields $H_{\a(2s+1)}$ and 
$H_{\a(2s)}$ defined according to \eqref{2.13}, and their gauge transformation
laws are given by eqs. \eqref{2.16} and \eqref{2.17}, respectively.
Now it is useful to split  each of $G_{\a(2s-2)} $ and $\O_{\a(2s-3)} $ into 
their real and imaginary parts, 
\bea
G_{\a(2s-2)}  = X_{\a(2s-2)}  +\ri Y_{\a(2s-2)} ~, \qquad
\O_{\a(2s-3)} =\F_{\a(2s-3)} +\ri \J_{\a(2s-3)} ~.
\eea
Then it follows from the gauge transformations \eqref{2.16},  \eqref{2.17}
and \eqref{4.6} that in fact we are dealing  with two different gauge theories. 
One of them is formulated in terms of the real unconstrained gauge superfields
\bea
\cV^{\parallel}_{s+\hf} = \big\{ H_{\a(2s+1)}, X_{\a(2s-2)}, \J_{\a(2s-3)}\}~,
\label{4.99}
\eea
which are defined modulo gauge transformations of the form
\begin{subequations} \label{4.100}
\bea
\d H_{\a_1 \dots \a_{2s+1} } 
&=& (-1)^s
\ri
D{}_{(\a_1}  \z_{\a_2 \dots \a_{2s+1})}~,\\
\d X_{\a_1 \dots \a_{2s-2} } &=&\frac{s^2}{2s+1} \pa^{\b\g} \z_{\b\g \a_1 \dots \a_{2s-2}} 
~,\\
\d \J_{\a_1 \dots \a_{2s-3}} &=& \frac{\ri s}{2(2s+1)} D^\b \pa^{\g\d} 
\z_{\b\g\d \a_1 \dots \a_{2s-3}} ~,
\eea
\end{subequations}
where the gauge parameter $\z_{\a(2s)}$ is real unconstrained. 
The other theory is described by the gauge superfields
\bea
\cV^{\parallel}_{s} = \big\{ H_{\a(2s)}, Y_{\a(2s-2)}, \F_{\a(2s-3)}\}~,
\label{4.111}
\eea
with the following gauge freedom
\begin{subequations} \label{4.122}
\bea
\d H_{\a_1 \dots \a_{2s} } 
&=& (-1)^s D{}_{(\a_1}  \z_{\a_2 \dots \a_{2s})}~,\\
\d Y_{\a_1 \dots \a_{2s-2} } &=&-\hf (-1)^s D^\b \z_{\b \a_1 \dots \a_{2s-2}} 
~,\\
\d \F_{\a_1 \dots \a_{2s-3}} &=& (-1)^s s  \pa^{\b \g} 
\z_{\b\g \a_1 \dots \a_{2s-3}} ~,
\eea
\end{subequations}
with the gauge parameter $\z_{\a(2s-1)}$ being real unconstrained. 


\subsection{Transverse formulation}

The transverse formulation is realised in terms of the following  dynamical variables:
\begin{align}
\mathcal{V}^{\perp}=&\big\{ {\mathbb H}_{\a(2s)},\, {\bf\G}_{\a(2s-2)},
\,\bar {\bf \G}_{\a(2s-2)} \big\} ~,
\label{4.12}
\end{align}
where the real superfield ${\mathbb H}_{\a(2s)} ={\mathbb H}_{(\a_{1}\dots \a_{2s})}$ is unconstrained, 
and the complex superfield ${\bf \G}_{\a(2s-2)}= {\bf \G}_{(\a_1 \dots \a_{2s-2})}$ is  
transverse linear,  
\bea
\bar{\mathbb D}^\b {\bf \G}_{\b \a_{1}\dots\a_{2s-3}}  =0\quad \Longrightarrow
\quad \bar{\mathbb D}^2 {\bf \G}_{\a(2s-2)}=0~.
\label{4.13}
\eea
The dynamical superfields are defined modulo 
gauge transformations of the form 
\begin{subequations} 
\label{4.14}
\bea
\delta {\mathbb H}_{\a_1 \dots \a_{2s}}&=& g_{\a_1 \dots \a_{2s}}
+\bar{g}_{\a_1 \dots \a_{2s}}~,  
\label{4.14a}
\\
\delta {\bf \Gamma}_{\a(2s-2)}&=&
\frac{s}{2s+1}\bar{D}^{\b}D^{\g}\bar{g}_{\a(2s-2)\b \g}~.
\label{4.14b}
\eea
\end{subequations}
The gauge-invariant action is 
\bea
S^{\perp}_{s+\frac{1}{2}}[{\mathbb H},{\bf \G}, \bar {\bf \G}]
&=& \Big(-\frac{1}{2}\Big)^{s}\int \rd^{3|4}z\,
\bigg\{\frac{1}{8}{\mathbb H}^{\a(2s)}D^{\b}\bar{D}^{2}D_{\b}
{\mathbb H}_{\a(2s)}
\nonumber \\
&&+{\mathbb H}^{\a(2s)}\left(D_{\a_1}\bar{D}_{\a_2} {\bf \G}_{\a_3 \dots \a_{2s}}
-\bar{D}_{\a_1} D_{\a_2}\bar{\bf \Gamma}_{\a_3\dots \a_{2s}}\right)\nonumber \\
&&+\frac{2s-1}{s}\bar{\bf \G}\cdot {\bf \G} 
+ \frac{2s+1}{2s}\left({\bf  \G} \cdot {\bf \G}+\bar{\bf \G}\cdot \bar {\bf \G}
\right)\bigg\}~.
\label{4166}
\eea

From the point of view of $\cN=1$ supersymmetry, 
${\bf \G}_{\a(2s-2)}$ is equivalent to two {\it complex unconstrained} superfields,
which we define as follows:
\bea
\G_{\a_1 \dots \a_{2s-2}}:={\bf \G}_{\a_1 \dots \a_{2s-2}}|~, \qquad
\U_{\a_1 \dots \a_{2s-1}}:= \ri  D^{\2}{}_ {(\a_1} {\bf \G}_{\a_2 \dots \a_{2s-1})}|~.
\label{4.15}
\eea
Making use of the gauge transformation \eqref{4.14b} gives 
\begin{subequations}
\bea
\delta {\bf \Gamma}_{\a(2s-2)}&=&
-\frac{\ri s}{2s+1}\pa^{\b\g}\bar{g}_{\a_1\dots \a_{2s-2}\b \g}
+
\frac{\ri s}{2s+1}{D}^{\1 \,\b}D^{\2\, \g}\bar{g}_{\a_1\dots \a_{2s-2}\b \g}~, \\
\ri  D^{\2}{}_ {(\a_1} \d {\bf \G}_{\a_2 \dots \a_{2s-1})} &=&
\frac{s}{2s+1} \Big\{ \ri \pa^\g{}_\b D^{\1 \, \b} \bar g_{\a_1 \dots \a_{2s-1} \g}
+\ri D^{\1\, \b} \pa^\g{}_{(\a_1} \bar g_{\a_2 \dots \a_{2s-1} )\b\g} \non \\
&&\quad + \pa^\b{}_{(\a_1} D^{\2 \, \g } \bar g_{\a_2 \dots \a_{2s-1} ) \b \g }
-\frac{\ri}{2} (D^{\1})^2 D^{\2\, \g } \bar g_{\a_1 \dots \a_{2s-1} \g}\Big\}~.~~~
\eea
\end{subequations}
From here we read off the gauge transformations of the $\cN=1$ superfields
\eqref{4.15}
\begin{subequations} \label{4.19}
\bea
\delta { \Gamma}_{\a(2s-2)}&=& \frac{s}{2(2s+1)}\pa^{\b\g}\z_{\a_1 \dots \a_{2s-2}\b \g}
-(-1)^s \frac{\ri}{2} D^\b \z_{\a_1 \dots \a_{2s-2} \b} ~, \\
\d \U_{\a(2s-1)}&=& - \frac{2}{2(2s+1)} \Big\{ 
\pa^\g{}_\b D^{ \b} \z_{\a_1 \dots \a_{2s-1} \g}
+D^{ \b} \pa^\g{}_{(\a_1} \z_{\a_2 \dots \a_{2s-1} )\b\g} \Big\} \non \\
&&\quad -\hf (-1)^s \Big\{ \pa^\b{}_{(\a_1}  \z_{\a_2 \dots \a_{2s-1} ) \b  }
-\frac{\ri}{2} D^2 \z_{\a_1 \dots \a_{2s-1} }\Big\}~.
\eea
\end{subequations}
Now it is useful to split  each of $\G_{\a(2s-2)} $ and $\U_{\a(2s-1)} $ into 
their real and imaginary parts, 
\bea
\G_{\a(2s-2)}  = X_{\a(2s-2)}  +\ri Y_{\a(2s-2)} ~, \qquad
\U_{\a(2s-1)} =\F_{\a(2s-1)} +\ri \J_{\a(2s-1)} ~.
\eea
Then it follows from the gauge transformations \eqref{2.16},  \eqref{2.17}
and \eqref{4.19} that in fact we are dealing  with two different gauge theories. 
One of them is formulated in terms of the real unconstrained gauge superfields
\bea
\cV^{\perp}_{s+\hf} = \big\{ H_{\a(2s+1)}, X_{\a(2s-2)}, \J_{\a(2s-1)}\}~,
\label{4.211}
\eea
which are defined modulo gauge transformations of the form
\begin{subequations} \label{4.222}
\bea
\d H_{\a_1 \dots \a_{2s+1} } 
&=& (-1)^s
\ri
D{}_{(\a_1}  \z_{\a_2 \dots \a_{2s+1})}~,\\
\d X_{\a_1 \dots \a_{2s-2} } &=&\frac{s}{2(2s+1)} \pa^{\b\g} \z_{\b\g \a_1 \dots \a_{2s-2}} 
~,\\
\d \J_{\a_1 \dots \a_{2s-1}} &=& \frac{\ri s}{2(2s+1)} 
 \Big\{ 
\pa^\g{}_\b D^{ \b} \z_{ \a_1 \dots \a_{2s-1} \g}
+D^{ \b} \pa^\g{}_{(\a_1} \z_{\a_2 \dots \a_{2s-1} )\b\g} \Big\} 
~,
\eea
\end{subequations}
where the gauge parameter $\z_{\a(2s)}$ is real unconstrained. 
The other theory is described by the gauge superfields
\bea
\cV^{\perp}_{s} = \big\{ H_{\a(2s)}, Y_{\a(2s-2)}, \F_{\a(2s-1)}\}~,
\label{4.233}
\eea
with the following gauge freedom
\begin{subequations}\label{4.244}
\bea
\d H_{\a_1 \dots \a_{2s} } 
&=& (-1)^s D{}_{(\a_1}  \z_{\a_2 \dots \a_{2s})}~,\\
\d Y_{\a_1 \dots \a_{2s-2} } &=&-\hf (-1)^s D^\b \z_{\b \a_1 \dots \a_{2s-2}} 
~,\\
\d \F_{\a_1 \dots \a_{2s-1}} &=& -\hf (-1)^s \Big\{ \pa^\b{}_{(\a_1}  \z_{\a_2 \dots \a_{2s-1} ) \b  }
-\frac{\ri}{2} D^2 \z_{\a_1 \dots \a_{2s-1} }\Big\}
~,
\eea
\end{subequations}
with the gauge parameter $\z_{\a(2s-1)}$ being real unconstrained. 


\subsection{Off-shell formulations for linearised $\cN=2$ supergravity}

The limiting case $s=1$ for the longitudinal and transverse formulations 
corresponds to linearised $\cN=2$ supergravity. As discussed in \cite{KO}, 
the longitudinal 
$s=1$ model is equivalent to the linearised action for type I minimal 
$\cN=2$ supergravity \cite{KT-M11}.
The transverse $s=1$ model is equivalent to the linearised action 
for $w=-1$ non-minimal $\cN=2$ supergravity \cite{KT-M11}.

\subsubsection{Longitudinal formulation: Type I minimal $\cN=2$ supergravity}
\label{typeI}

In the  $s=1$ case, 
the constraint \eqref{4.1} means that the $\cN=2$ superfield $\mathbb G$ is chiral. 
The specific feature of $s=1$ is that 
the second $\cN=1$ superfield in \eqref{4.4} does not exist in this case. 
According to \eqref{4.6}, the scalar $G={\mathbb G} |$ transforms as
\bea
\d G = \frac{1}{3} \pa^{\a\b} \z_{\a\b} +\frac{\ri}{2} D^\a \z_\a~.
\eea
We introduce the real and imaginary parts of $G$, $G= X +\ri Y$. 
From the point of view of $\cN=1$ supersymmetry, the original $\cN=2$ theory is 
equivalent to a sum of two models. One of them realises an 
 off-shell $\cN=1$ supergravity multiplet. It  is described by the real gauge fields
 \bea
\cV^{\parallel}_{3/2} = \big\{ H_{\a \b \g}, X\}~,
\eea
with the following gauge transformation law: 
\bea
\d H_{\a\b\g} = -\ri D_{(\a} \z_{\b\g)}~, \qquad 
\d X =  \frac{1}{3} \pa^{\a\b} \z_{\a\b} ~.
\label{4.27}
\eea
The second model realises an off-shell $\cN=1$ gravitino multiplet. 
It  is described by the real gauge fields
 \bea
\cV^{\parallel}_{1} = \big\{ H_{\a \b }, Y \}~,
\eea
with the following gauge transformation laws: 
\bea
\d H_{\a\b} = - D_{(\a} \z_{\b )}~, \qquad 
\d Y =  \hf D^{\a} \z_{\a} ~.
\eea


\subsubsection{Transverse formulation: Non-minimal $\cN=2$ supergravity}
\label{subsubsection4.3.2}

For $s=1$ the transverse linear constraint \eqref{4.13} is not defined. 
However, its corollary $ \bar{\mathbb D}^2 {\bf \G}_{\a(2s-2)}=0$ can be used;
for $s=1$ it defines a complex linear superfield. Then the  gauge-invariant action 
\eqref{4166} corresponds  to the linearised action 
for $w=-1$ non-minimal $\cN=2$ supergravity \cite{KT-M11}.
Upon reduction to $\cN=1$ superspace, this dynamical system describes
two off-shell $\cN=1$ supermultiplets, a supergravity multiplet and a gravitino multiplet.
The supergravity multiplet is described by the real gauge superfields
\bea
\cV^{\perp}_{3/2} = \big\{ H_{\a \b\g}, X, \J_{\a}\}~,
\eea
The gravitino multiplet is described by the real gauge superfields
\bea
\cV^{\perp}_{1} = \big\{ H_{\a \b}, Y, \F_{\a}\}~,
\eea


\subsubsection{Type II minimal $\cN=2$ supergravity}\label{typeII}

The linearised action for type II minimal $\cN=2$ supergravity \cite{KT-M11} is
\bea
S^{(II)}[{\mathbb H},{\mathbb S}]&=&
\int \rd^{3|4}z \,
\Big\{
-\frac{1}{16}{\mathbb H}^{\a\b}D^\g\bar D^2 D_\g {\mathbb H}_{\a\b}
-\frac{1}{4}(\pa_{\a\b}{\mathbb H}^{\a\b})^2
+\frac{1}{16}([D_\a,\bar D_\b] {\mathbb H}^{\a\b})^2
\non\\
&&~~
+\frac{1}{4}{\mathbb S}[D_\a,\bar D_\b] {\mathbb H}^{\a\b}
+ \hf {\mathbb S}^2\Big\} ~,
\label{II-3D-H}
\eea
where the real compensator $\bar {\mathbb S} = {\mathbb S}$ is a linear superfield, 
\bea
\bar {\mathbb D}^2  {\mathbb S}={\mathbb D}^2  {\mathbb S}=0~.
\label{4.31}
\eea
Such a superfield describes the field strength of an Abelian $\cN=2$ vector multiplet.
The action \eqref{II-3D-H} is invariant under the gauge transformations
\bea
\d {\mathbb H}_{\a\b} = g_{\a\b} +\bar g_{\a\b}~, \qquad 
\d {\mathbb S} = -\frac{1}{3} ({\mathbb D}^\a \bar{\mathbb D}^\b g_{\a\b} 
- \bar{\mathbb D}^\a {\mathbb D}^\b \bar g_{\a\b} )~.
\label{4.32}
\eea

The linear constraint \eqref{4.31} is equivalent to two constraints in the real basis for the 
covariant derivatives. The constraints are 
\begin{subequations}
\bea
\Big\{ (D^{\2})^2 - (D^{\1})^2 \Big\} {\mathbb S} &=&0~,\\
D^{\1\, \a } D^{\2}{}_\a {\mathbb S}&=&0~.
\eea
\end{subequations}
Thus $\mathbb S$ is equivalent to the following real $\cN=1$ superfields
\bea
X:= {\mathbb S} |~, \qquad W_\a := -\ri D^{\2 }{}_\a {\mathbb S}|~,
\label{S-components}
\eea
of which the former is unconstrained and the latter is the field strength of an Abelian 
$\cN=1$ vector multiplet (see, e.g., \cite{GGRS}), 
\bea
D^\a W_\a =0~.
\label{N=1WMBI}
\eea
To derive the gauge transformations of $X$ and $W_\a$, we should rewrite 
 the gauge transformation of $\mathbb S$, eq.  \eqref{4.32}, 
as well as its corollary $D^{\2}{}_\a \d {\mathbb S}$, 
 in the real basis for the covariant derivatives. We obtain 
 \begin{subequations}
 \bea
 \d {\mathbb S} &=& \frac{\ri}{3} \pa^{\a\b} (g_{\a\b} - \bar g_{\a\b} ) 
 +\frac{\ri}{3} D^{\1\,\a} ( D^{\2 \, \b} g_{\a\b} + D^{\2 \, \b } \bar g_{\a\b})~,\\
 -\ri D^{\2 }{}_\a \d {\mathbb S} &=& - \frac{\ri}{3} \Big(
 D^\b \pa_\a{}^\g g_{\b\g} 
+\pa^\b{}_\g D^\g g_{\a\b}
+\ri \pa_\a{}^\b D^{\2\, \g} g_{\b \g} 
\non \\
&& \qquad \qquad \qquad \qquad 
-\hf
(D^{\1})^2 D^{\2\, \b} g_{\a\b}\Big) +{\rm c.c.}~~~
 \eea 
 \end{subequations}

From the point of view of $\cN=1$ supersymmetry, 
the dynamical system under consideration splits into two $\cN=1$ supersymmetric 
theories. One of them describes the off-shell $\cN=1$ supergravity multiplet 
realised in terms of the gauge superfields
 \bea
\cV^{(II)}_{3/2} = \big\{ H_{\a \b \g}, X\}~,
\eea
with the gauge transformation of $X$ being identical to that of $X$ in \eqref{4.27}.
The other provides an off-shell realisation for $\cN=1$ gravitino multiplet 
realised in terms of the gauge superfields
 \bea
\cV^{(II)}_{1} = \big\{ H_{\a \b }, W_\a\}~.
\eea
Their gauge transformation laws are:
\bea
\d H_{\a\b} = - D_{(\a} \z_{\b )}~, \qquad 
\d W_\a = \ri D^\b D_\a \z_\b~.
\eea


\subsubsection{Type III minimal $\cN=2$ supergravity}\label{typeIII}

Type III supergravity \cite{KT-M11} is described by the action 
\bea
S^{(III)}  [ {\mathbb H},\mathbb T] &=& \int \rd^{3|4} z \,
\Big\{
-\frac{1}{16} {\mathbb H}^{\a\b}D^\g\bar D^2 D_\g {\mathbb H}_{\a\b}
-\frac{1}{8}(\pa_{\a\b} {\mathbb H}^{\a\b})^2
+\frac{1}{32}([D_\a,\bar D_\b]  {\mathbb H}^{\a\b})^2
\non\\
&&~~
+\frac{1}{4}{\mathbb{T}}\pa_{\a\b}  {\mathbb H}^{\a\b}
+\frac{1}{8}  {\mathbb{T}}^2
\Big\}~,
\label{III}
\eea
where the real compensator $\bar {\mathbb T} = {\mathbb T}$ is a linear superfield, 
\bea
\bar {\mathbb D}^2  {\mathbb T}={\mathbb D}^2  {\mathbb T}=0~.
\eea
The action \eqref{III} is invariant under the gauge transformations
\bea
\d {\mathbb H}_{\a\b} = g_{\a\b} +\bar g_{\a\b}~, \qquad 
\d {\mathbb T} = -\frac{\ri}{3} (
{\mathbb D}^\a \bar{\mathbb D}^\b g_{\a\b} 
+ \bar{\mathbb D}^\a {\mathbb D}^\b \bar g_{\a\b} )~
\eea
compare with \eqref{4.32}

The compensator  $\mathbb T$ is equivalent to the following real $\cN=1$ superfields
\bea
T:= {\mathbb T} |~, \qquad Z_\a := -\ri D^{\2 }{}_\a {\mathbb T}|~,
\eea
of which the former is unconstrained and the latter is the field strength of an Abelian 
$\cN=1$ vector multiplet,
\bea
D^\a Z_\a =0~.
\eea
Upon reduction to $\cN=1$ superspace, the theory describes two off-shell $\cN=1$ 
supermultiplets, a supergravity multiplet and an gravitino multiplet. 
The supergravity multiplet is realised in terms of the gauge superfields 
 \bea
\cV^{(III)}_{3/2} = \big\{ H_{\a \b \g}, Z_\a \}~.
\eea
Their gauge transformations are:
\bea
\d H_{\a\b\g} = -\ri D_{(\a} \z_{\b\g)}~, \qquad 
\d Z_\a = - \frac{1}{3} D^\b D_\a D^\g \z_{\b\g} ~.
\eea
The gravitino multiplet 
is realised in terms of the gauge superfields
 \bea
\cV^{(III)}_{1} = \big\{ H_{\a \b }, T\}~,
\eea
with the gauge freedom
\bea
\d H_{\a\b} = - D_{(\a} \z_{\b )}~, \qquad 
\d T =  D^\a \z_\a ~.
\eea


\section{Off-shell formulations for massless low spin supermultiplets}
\label{section5}

Consider an arbitrary $\cN=2$ supersymmetric theory with action 
\bea
S =\int \rd^{3|4}z \, L_{(\cN=2)}~,
\eea
where the Lagrangian $L_{(\cN=2)}$ is a real scalar $\cN=2$ superfield.
The action can be reduced to component fields by the rule
\bea
S = \int \rd^3 x \, L_{(\cN=0)}~, \qquad 
L_{(\cN=0)}: = \frac{1}{16} {\mathbb D}^2 \bar{\mathbb D}^2L_{(\cN=2)}\Big|_{\q_I =0}
~,
\eea
or to $\cN=1$ superspace
\bea
S = \int \rd^{3|2} z \, L_{(\cN=1)}~, \qquad 
L_{(\cN=1)} := -\frac{\ri}{4} (D^{\2})^2 L_{(\cN=2)}\Big|~.
\eea
Here the  Lagrangian $L_{(\cN=1)}$ is a real scalar $\cN=1$ superfield.

\subsection{Scalar and vector multiplets}

As an example, we consider the low-energy model for an Abelian 
$\cN=2$ vector multiplet with Lagrangian $ L_{(\cN=2)} = \cF ({\mathbb S})$, 
where the vector multiplet field strength $\mathbb S$  is a real linear superfield
constrained as in \eqref{4.31}. Upon reduction to $\cN=1$ superspace, the action becomes
\bea
S &=& \frac{\ri}{4} \int \rd^{3|2} z \, \Big\{ 
  \cF'' (X) W^\a W_\a - \cF' (X) D^2 X
\Big\} 
\non \\
&=& \frac{\ri}{4} \int \rd^{3|2} z \, \cF'' (X) \Big\{ D^\a X D_\a X + W^\a W_\a\Big\} ~,
\label{5.5}
\eea
where the $\cN=1$ components of $\mathbb S$, 
$X$ and $W_\a$,  are defined as in \eqref{S-components}.
We recall that the $\cN=1$ field strength $W_\a$ obeys the 
Bianchi identity \eqref{N=1WMBI}, 
 which is solved according to \eqref{3.36}.
For a free $\cN=2$ vector multiplet, $\cF ({\mathbb S}) = - {\mathbb S}^2$.

The model for a massless $\cN=1$ scalar multiplet is
\bea
S_{\rm SM} =
-\frac{\ri}{2} \int \rd^{3|2} z \,  D^\a X D_\a X~.
\label{5.6}
\eea
The model for a massless $\cN=1$ vector multiplet is 
\bea
S_{\rm VM} &=&- \frac{\ri}{2} \int \rd^{3|2} z \,  W^\a W_\a 
= \frac{1}{4}\int \rd^{3|2} z \,\Big\{ -\ri H^\a \Box H_\a + \hf H^\a \pa_{\a\b} D^2 H^\b\Big\}~.
\label{5.7}
\eea
Here we have made use of  the gauge prepotential $H_\a$ for the vector  
multiplet, eq. \eqref{3.36}.
The models $S_{\rm SM} $ and $S_{\rm VM} $ are dual to each other
\cite{HKLR}.
It is worth reviewing this duality, for we will meet other examples of dual $\cN=1$ 
supersymmetric field theories. Consider a model for the vector multiplet $W_\a$ coupled to a background superfield $\L$ with action 
\bea
S =- \frac{\ri}{2} \int \rd^{3|2} z \,  \L W^\a W_\a~, \qquad D^\a W_\a =0~.
\label{5.8}
\eea
This model is equivalent to a first-order model with action 
\bea
S_{\mbox{\footnotesize{first-order}}} =- \frac{\ri}{2} \int \rd^{3|2} z \,   \L \cW^\a \cW_\a +  \int \rd^{3|2} z \,   \cW^\a D_\a X~,
\label{5.9}
\eea
in which the dynamical variables are  an unconstrained real spinor superfield
$\cW_\a$ and a real scalar superfield $X$. Varying $X$ gives $D^\a \cW_\a =0$, 
and hence $\cW_\a = W_\a$. As a result, the action \eqref{5.9} reduces to 
\eqref{5.8}. On the other hand, we can integrate out the auxiliary superfield $\cW_\a$ 
from \eqref{5.9} to result with the dual action 
\bea
S_{(dual)}= - \frac{\ri}{2} \int \rd^{3|2} z \,   \L^{-1}  D^\a X D_\a X~.
\label{5.10}
\eea
The inverse duality transformation is obtained by replacing \eqref{5.10} 
with an equivalent first-order action 
\bea
\widetilde{S}_{\mbox{\footnotesize{first-order}}}= - \frac{\ri}{2} \int \rd^{3|2} z \,   \L^{-1}  \cY^\a \cY_\a  
 +  \int \rd^{3|2} z \,   W^\a \cY_\a ~,
\eea
in which $\cY_\a$ is an unconstrained {\it imaginary} spinor superfield, 
and $W_\a$ the field strength of a vector multiplet. 


\subsection{Gravitino multiplet} \label{section5.2}

An off-shell formulation for the massless $\cN=1$ gravitino multiplet 
can be realised in terms of two real unconstrained gauge 
superfields, a three-vector 
$H_{\alpha \beta} = H_{\b\a}$ and a scalar   $X$.
The  superfield Lagrangian has the form
\be 
\label{LFL}
L_{\rm GM} = -\frac{\ri}{2} \Big\{  H^{\alpha \beta} D^2 H_{\alpha \beta}
+ D_\alpha H^{\alpha \beta} D_\gamma H^{\gamma}{}_{\beta}
- 2  D_{\a} H^{\alpha \beta} D_\b X
- D^\b X D_\b X \Big\}~,
\ee
and proves to be equivalent to the Lagrangian introduced by Siegel \cite{Siegel}.
The action associated with  \p{LFL} is invariant under the gauge transformations
\be 
\label{LFG}
\delta H_{\alpha \beta} = D_{\alpha} \z_{\beta } + D_{\beta} \z_{\alpha }
=2D_{(\a} \z_{\b)}~, 
\qquad \delta X = D^\alpha  \z_\alpha
\ee
where the gauge parameter $\z_\alpha$ is 
real unconstrained. The superfield $X$ is a compensator, 
for its kinetic term in \eqref{LFL} has a  wrong sign as 
compared with the scalar multiplet  \eqref{5.6}.

The gravitino multiplet possesses a dual formulation obtained by dualising the 
scalar compensator in \eqref{LFL} into a vector multiplet. 
The dual Lagrangian is 
\bea
L^{(dual)}_{\rm GM} =- \frac{\ri}{2} \Big\{   H^{\alpha \beta} D^2 H_{\alpha \beta} 
+ 2D_\alpha H^{\alpha \beta} D_\gamma H^{\gamma}{}_{\beta} 
+ 2\ri W^\a D^\b H_{\a\b}  - W^\a W_\a \Big\}~.~~
\eea
It is invariant under the gauge transformations 
\bea 
\delta H_{\alpha \beta} = D_{\alpha} \z_{\beta } + D_{\beta} \z_{\alpha }~, 
\qquad \delta W_\a = 2{\ri} D^\b D_\a \z_\b ~.
\eea

\subsection{Supergravity multiplet}\label{LONGI}

An off-shell formulation for the massless $\cN=1$ supergravity multiplet 
can be realised in terms of two real unconstrained gauge 
superfields, a symmetric rank-3 spinor 
$H_{\alpha \beta \gamma} =H_{(\a\b\g)}$ and 
a scalar $X$.
The superfield Lagrangian is 
\bea \nonumber
L_{\rm SGM} &=& 
\frac{\ri}{4}  H^{\alpha \beta \gamma} \Box H_{\alpha \beta \gamma}
- \frac{1}{8}H^{\alpha \beta \gamma} \partial_{\gamma \rho} D^2
H_{\alpha \beta}{}^{ \rho}
- \frac{\ri}{4} 
\partial_{\a \b}H^{\a \b \gamma} \partial^{\rho \sigma}H_{\rho \sigma}{}_{\gamma} \\
&& +\hf \partial_{\a\b} H^{\a \b \gamma}D_\gamma X 
+ \frac{\ri}{2} D^\g X D_\g  X~.
 \label{LHL}
\eea
The Lagrangian \p{LHL} is invariant under the gauge transformations \cite{GGRS}
\begin{equation} \label{LHG}
\delta H_{\alpha \beta \gamma} = \ri ( D_\alpha \z_{\beta \gamma} + 
D_\beta \z_{ \alpha \gamma }+D_\gamma \z_{\alpha \beta })
=3\ri D_{(\a}\z_{\b\g)}~, 
\qquad \delta X = - \partial^{\alpha \beta} \z_{\alpha \beta}~.
\end{equation}

The supergravity multiplet possesses a dual formulation obtained by dualising the 
scalar compensator in \eqref{LHL} into a vector multiplet. 
The dual Lagrangian is 
\bea \nonumber \
L^{(dual)}_{\rm SGM}&=& \frac{\ri}{4}  H^{\alpha \beta \gamma} \Box H_{\alpha \beta \gamma}
- \frac{1}{8}H^{\alpha \beta \gamma} \partial_{\gamma \rho} D^2
H_{\alpha \beta}{}^{ \rho}
- \frac{\ri}{8} 
\partial_{\a \b}H^{\a \b \gamma} \partial^{\rho \sigma}H_{\rho \sigma}{}_{\gamma} \\
&& +\frac{\ri}{2} \partial_{\a\b} H^{\a \b \gamma}W_\g
+ \frac{\ri}{2} W^\a W_\a~.
\eea
It is invariant under gauge transformations 
\bea
\delta H_{\alpha \beta \gamma} = \ri ( D_\alpha \z_{\beta \gamma} + 
D_\beta \z_{ \alpha \gamma }+D_\gamma \z_{\alpha \beta })~, \qquad
\d W_\a = \hf D^\b D_\a D^\g \z_{\b\g}~.
\eea

At this stage it is worth pausing in order to discuss some of the results obtained. 
According to the classification of linearised off-shell actions for 
4D $\cN=1$ supergravity  \cite{GKP}, there are three minimal 
models with $12+12$ off-shell degrees of freedom. The three-dimensional $\cN=2$ analogues of these models (with $8+8$ off-shell degrees of freedom) 
were constructed in \cite{KT-M11} 
and called the type I, type II and type III minimal supergravity theories. 
We discussed these models in sections \ref{typeI}, \ref{typeII} and \ref{typeIII},
respectively. The difference between the 3D $\cN=2$ minimal supergravity models becomes quite transparent upon their reduction to $\cN=1$ superspace. 
Every  $\cN=2$ action becomes a sum of two $\cN=1$ actions, one of which describes 
the  gravitino multiplet and the other corresponds to the supergravity multiplet. 
Each of the $\cN=1$ actions is realised in terms of two $\cN=1$ superfields, 
of which one is universally the superconformal gauge field ($H_{\a\b}$ for the gravitino 
multiplet, $H_{\a\b\g}$ for the supergravity multiplet), while the other is a compensator. 
The difference between the three minimal $\cN=2$ supergravity models is encoded 
in different types of $\cN=1$ compensators. In the case of type I supergravity, 
both the $\cN=1$  supergravity and gravitino multiplets are characterised by scalar compensators, devoted $X$ and $Y$, respectively, in  section \ref{typeI}.
The type II and type III formulations are obtained by dualising one 
of the scalar $X$ and $Y$ into an $\cN=1$  vector multiplet.
In principle, it is possible to dualise both $X$ and $Y$ into vector multiplets. 
This would lead to a new linearised action for $\cN=2$ supergravity involving 
a double vector multiplet \cite{IKL} as the corresponding $\cN=2$ compensator. 
However, such an action proves to possess $\cN=2$ supersymmetry with an intrinsic central charge, which is less interesting than the standard $\cN=2$ Poincar\'e supersymmetry.


\subsection{Transverse formulation}

The models \eqref{LFL} and \eqref{LHL} correspond to the longitudinal formulation
discussed in section \ref{typeI}. It is of interest to compare them with the models
originating within the transverse formulation sketched in section 
\ref{subsubsection4.3.2}.

In addition to the dynamical variables $H_{\a\b}$ and $X$, the gravitino multiplet
now contains an auxiliary spinor superfield $\F_\a$. 
The  Lagrangian has the form
\bea \label{TFL}
L_{\rm GM}^{\perp} &=& -\frac{\ri}{2}  H^{\alpha \beta} D^2 H_{\alpha \beta} 
- \frac{\ri}{2}  D_\a H^{\alpha \beta} D_\b X
-\hf \F^\a D^\b  H_{\alpha \beta} \non\\
&&+ \frac{\ri}{4} D^\a X D_\a X - \frac{\ri}{2} \F^\alpha \F_\alpha 
-\hf \F^\a  D_\alpha X~.
\eea
The Lagrangian \p{TFL} is invariant under the gauge transformations
\bea \label{TFG}
&&\delta H_{\alpha \beta} = D_{\alpha} \z_{\beta } + D_{\beta} \z_{\alpha }~,\quad 
 \delta X = D^\alpha  \z_\alpha~, \quad
 \delta \F_\alpha =\partial_{\alpha \beta} \z^\beta + \frac{\ri}{2} D^2 \z_\alpha~.
\eea
The superfield $\F_\a$ can be integrated out 
using its equation of motion 
\be
\F^\alpha = - \frac{\ri}{2} D_\beta H^{\alpha \beta} + \frac{\ri}{2} D^\alpha X~.
\ee
Then the Lagrangian \p{TFL} reduces to  \p{LFL}.

Within the transverse formulation, 
the supergravity multiplet contains not only the dynamical variables
  $H_{\alpha \beta \gamma} $ and $X$, but also an auxiliary spinor superfield
$\J_{\alpha}$.
The corresponding Lagrangian is
\bea  \label{THL} \nonumber
L&=& \frac{\ri}{4} H^{\alpha \beta \gamma} \Box H_{\alpha \beta \gamma}
- \frac{1}{8}H^{\alpha \beta \gamma} D^2\partial_{\gamma \rho} H_{\alpha \beta}{}^\r
-\hf  \partial_{\alpha \beta} H^{\alpha \beta \gamma}  (\ri \J_\gamma + D_\g X) \\
&& 
+\frac{\ri}{4} \J^\a \J_\a +\J^\a D_\a X
-\frac{\ri}{2} D^\a X D_\a  X ~.
\eea
The action associate with this Lagrangian is invariant 
under the gauge transformations
\begin{subequations}\label{THG}
\begin{eqnarray}
&&\delta H_{\alpha \beta \gamma} = \ri( D_\alpha \z_{\beta \gamma} + 
D_\beta \z_{ \alpha \gamma }+D_\gamma \z_{\alpha \beta })~, \\ 
&&\delta \J_\alpha = \ri( \partial^{\b \g} D_\b \z_{\alpha \g} 
- D_\b \partial_{\alpha \g} \z^{\b\g}) ~, \\
&&\delta X = - \ri \partial^{\alpha \beta} \z_{\alpha \beta}~.
\end{eqnarray}
\end{subequations}
The auxiliary superfield $\J_\a$ can be integrated out using its equation of motion
\be
 \J^\alpha = \partial_{\b \g} H^{\alpha \b\g } + 2 \ri D^\alpha  X~.
\ee
Then the Lagrangian \p{THL} turns into 
\p{LHL}.


\section{Massless higher spin supermultiplets}\label{section6}

To derive off-shell formulations for the massless higher spin $\cN=1$ supermultiplets, 
one can apply the $\cN=2 \to \cN=1$ superspace reduction to the actions
\eqref{N2LongAction} and \eqref{4166}.  Here we will follow a different approach. 
We make use of the two pieces of input information: 
(i) the four sets of dynamical variables 
$\cV^{\parallel}_{s+\hf} $, $\cV^{\parallel}_{s}$,  $\cV^{\perp}_{s+\hf}$ and 
$\cV^{\perp}_{s}$ defined by eqs. \eqref{4.99} \eqref{4.111}, \eqref{4.211} and
\eqref{4.233}, respectively; and (ii) the corresponding gauge transformation laws given
by eqs. \eqref{4.100}, \eqref{4.122}, \eqref{4.222} and \eqref{4.244}, respectively.  
To construct gauge-invariant actions, we will make use of the oscillator realisation for higher spin fields, see \cite{FT} for a review.\footnote{The oscillator formulation 
for the off-shell massless higher spin $\cN=1$ supermultiplets in four dimensions \cite{KSP,KS93} was presented in \cite{GK}.}
The oscillator construction 
is expected to be useful for deriving interaction vertices 
for higher spin supermultiplets. 

Before we proceed, a comment on the terminology used below is in order. 
In three dimensions, the notion of superspin is defined only in the massive case, 
see section \ref{section2.2}. When speaking of a massless higher superspin theory
in three dimensions, we will refer to the kinematic structure of the field variables, 
their gauge transformation laws and the gauge-invariant action. 
Given an  integer $s>1$, the massless supersymmetric gauge theories described by
the dynamical variables $\cV^{\parallel}_{s}$ or $\cV^{\perp}_{s}$ 
will be referred to as massless integer superspin multiplets, 
for the gauge superfield $H_{\a(2s)}$ carries an even number of spinor indices. 
When speaking of massless half-integer superspin multiplets, 
we mean 
the massless supersymmetric gauge theories described by
the dynamical variables $\cV^{\parallel}_{s +\hf }$ or $\cV^{\perp}_{s+\hf }$, 
for the gauge superfield $H_{\a(2s+1)}$ carries an odd number of spinor indices. 


\subsection{Auxiliary oscillators} \label{OSCILLATORS}

In order to simplify computations for higher spin superfields, let us introduce 
auxiliary oscillators defined by the commutation relations
\be
[ a^\alpha, a^{\beta +} ]= \varepsilon^{\ga \gb}~.
\ee
An ``$n$-particle''  ket-state $| \Phi_n \rangle$ in this auxiliary Fock space is defined as
\be \label{FS}
| \Phi_n \rangle = \frac{1}{n!} \Phi_{\ga_1 \ga_2 \dots \ga_n} (z) a^{\ga_1+ }  a^{\ga_2+ }\dots a^{\ga_n+ }
| 0\rangle~, 
\ee
with the Fock vacuum defined by $a^\ga |0\rangle=0$. Here $\F_{\a(n)} (z)$ 
is a symmetric rank-$n$ spinor superfield. 
The bra-state $\langle \Phi_n |$ is defined similarly, 
\bea
\langle \Phi_n | = \frac{1}{n!} \langle 0| a^{\ga_1 }  a^{\ga_2 }\dots a^{\ga_n }
 \Phi_{\ga_1 \ga_2 \dots \ga_n}  (z) ~.
\eea
We introduce the following operators
\begin{subequations}
\bea \label{op1}
\gamma &=& a^\ga D_ \ga~, \quad \gamma^+ = a^{\ga +} D_ \ga~, \\
\label{op3}
P &=& a^\alpha \partial_{\alpha \beta} D^\beta~, \quad P^+ = a^{\alpha+} \partial_{\alpha \beta} D^\beta~, 
\\
\label{op2}
K_l &=& a^{\ga_1+ }  
\dots a^{\ga_l+ } \pl_{\ga_1 \gb_1} 
\dots \pl_{\ga_l \gb_l}
a^{\gb_1 }  
\dots a^{\gb_l }~.
\eea
\end{subequations}
Some properties of these operators are listed in Appendix \ref{Appendix C}.

The action of the operators \p{op1}--\p{op2} on a state of the form  \p{FS} can be translated as follows
\begin{subequations}
\bea
\gamma | \Phi_n \rangle & \rightarrow &D_{\b} \Phi^{\b}{}_{ \ga_1\dots \ga_{n-1}}~, \quad
\gamma^+ | \Phi_n \rangle \rightarrow (n+1)D_{(\ga_1} \Phi_{\ga_2  \dots \ga_{n+1})}~,\\
P| \Phi_n \rangle & \rightarrow & D^{\gb} \pl_{\gb \g}\Phi^{\g}{}_{ \ga_1 \dots \ga_{n-1}}~, \quad
P^+ | \Phi_n \rangle \rightarrow (n+1) D^{\gb} \pl_{\gb (\ga_1}\Phi_{ \ga_2  \dots \ga_{n+1})}~,\\
K_l  | \Phi_n \rangle &\rightarrow & 
(-1)^l \frac{n!}{(n-l)!} \pa^{\b_1}{}_{(\a_1} \dots \pa^{\b_l}{}_{\a_l} 
\F_{\a_{l+1} \dots \a_n ) \b_\a \dots \b_l} 
~.
\eea
\end{subequations}
We also introduce the ``number operator"  $N= a^{\alpha +} a_\alpha$ 
which acts on $|\Phi_n \rangle$ as
\be
 N | \Phi_n \rangle = n  | \Phi_n \rangle~.
\ee

\subsection{Integer superspin multiplets} \label{His-l}

In this and the next subsections, we present massless gauge theories
realised in terms of the  dynamical variables $\cV^{\parallel}_{s}$ and
$\cV^{\parallel}_{s+\hf} $  defined by eqs. \eqref{4.111} and 
\eqref{4.99}, respectively.

A Lagrangian formulation for a massless multiplet of integer superspin $s$, with $s>1$, 
contains a gauge superfield $| H_{2s} \rangle$, a compensator $| Y_{2s-2} \rangle$ 
and an auxiliary superfield $|\F_{2s-3} \rangle$.
The superfield Lagrangian, ${\cal L}^{\parallel}_s$, is
\bea 
\frac{(-1)^s}{(2s-1)!}  {\cal L}^{\parallel}_s&=&\frac{\ri}{2} \langle H_{2s} | N D^2 | H_{2s} \rangle 
-\frac{\ri}{2}\langle H_{2s} | \gamma^+ \gamma | H_{2s} \rangle  \non \\
&&-\frac{\ri}{2}\langle {Y}_{2s-2} | \g^2 | H_{2s} \rangle  
+\frac{\ri}{2}  \langle H_{2s} | \g^{+2} | {Y }_{2s-2} \rangle  
+\frac{\ri}{2}\langle {Y}_{2s-2} | D^2 | {Y}_{2s-2} \rangle
\non \\ 
&& 
-\langle \F_{2s-3} | \gamma | Y_{2s-2} \rangle 
- \langle Y_{2s-2} | \gamma^+ | \F_{2s-3} \rangle
+ 2\ri \langle \F_{2s-3} | | \F_{2s-3} \rangle ~.
\label{L1-1}
\eea
The corresponding action 
proves to be  invariant under gauge transformations 
\begin{subequations}\label{GT1-1}
\bea 
&& \delta | H_{2s} \rangle =  \gamma^+ | \z_{2s-1} \rangle ~,\\ 
&& \delta | Y_{2s-2} \rangle =  \gamma | \z_{2s-1}\rangle ~,\\ 
&& \delta | \F_{2s-3} \rangle = -\frac{\ri}{2} \g^2 | \z_{2s-1}\rangle ~.
\eea
\end{subequations}
The equation of motion for $| \F_{2s-3} \rangle$ 
expresses this field in terms of $| Y_{2s-2} \rangle$,
\be
| \F_{2s-3} \rangle = -\frac{\ri}{2}\gamma | Y_{2s-2} \rangle~.
\ee
Plugging 
this expression back into the Lagrangian \p{L1-1} gives
\bea 
\frac{(-1)^s}{(2s-1)!} {\cal L}_{s}&=& \frac{\ri}{2} \langle H_{2s} |N D^2 | H_{2s} \rangle 
-\frac{\ri}{2}\langle H_{2s} | \gamma^+ \gamma | H_{2s} \rangle 
-\frac{\ri}{2} 
\langle Y_{2s-2} | \g^2 | H_{2s} \rangle \non  \\ 
&&
 + \frac{\ri}{2} \langle H_{2s} | \g^{+2}| Y_{2s-2} \rangle  
+\frac{\ri}{2}\langle Y_{2s-2} | D^2 | Y_{2s-2} \rangle +
 \frac{\ri}{2}\langle Y_{2s-2} | \gamma^+ \gamma| Y_{2s-2} \rangle ~.~~~~~
\label{L1-2}
\eea

The above results can  be  readily recast in terms of ordinary superfields.
We introduce the gauge superfields $H_{\a(2s)}$, $Y_{\a(2s-2)}$ 
and $\F_{\a(2s-3)} $ as follows:
\begin{subequations}
\bea
&&| H_{2s} \rangle = \frac{1}{(2s)!} H_{\ga_1 \dots \ga_{2s}} a^{\ga_1+ }  
\dots a^{\ga_{2s}+ } | 0\rangle~, 
\\ 
&&|Y_{2s-2} \rangle = \frac{1}{(2s-2)!} Y_{\ga_1 \dots \ga_{2s-2}} a^{\ga_1+ }  
\dots a^{\ga_{2s-2}+ } | 0\rangle~, \\ 
&&|\F_{2s-3} \rangle = \frac{1}{(2s-3)!} 
\F_{\ga_1 \dots \ga_{2s-3}} a^{\ga_1+ } \dots a^{\ga_{2s-3}+ } | 0\rangle~.
\eea
\end{subequations}
The gauge parameters $\z_{\a(s2-1)}$ are introduced similarly, 
\be
|\z_{2s-1}  \rangle = \frac{1}{(2s-1)!} \z_{\ga_1 \dots \ga_{2s-1}} a^{\ga_1+ }  \dots a^{\ga_{2s-1}+ } | 0\rangle ~.
\ee
 The Lagrangian \p{L1-1} is equivalently written as 
\bea 
(-1)^s {\cal L}^{\parallel}_s &=& \frac{\ri}{2} H_{\ga_1 \dots\ga_{2s}} D^2 H^{\ga_1\dots \ga_{2s}} 
+ \frac{\ri}{2} D^\b H_{\b \ga_1 \dots  \ga_{2s-1} } D_\g H^{\g \ga_1 \dots  \ga_{2s-1} } \non \\ 
\nonumber
&&+
(2s-1) Y_{\ga_1 \dots \ga_{2s-2}  } \pl_{\b \g } 
H^{ \b \g \ga_1 \dots \ga_{2n-2} } 
+\frac{\ri}{2}  (2s-1) {Y}_{\ga_1 \dots \ga_{2s-2}} D^2 {Y}^{\ga_1 \dots \ga_{2s-2}}  \\ 
\nonumber
&&+2 \ri  (2s-1) (2s-2)\F_{\ga_1 \dots \ga_{2s-3}} \F^{\ga_1 \dots \ga_{2s-3}} 
 \\ 
&&-2 (2s-1) (2s-2) \F_{\ga_1 \dots \ga_{2s-3}} 
D_{\b} {Y }^{\b \ga_1 \dots \ga_{2s-3}}~,
\label{L1-3-0}
\eea
while the Lagrangian \p{L1-2} coincides with
\bea
(-1)^s {\cal L}_{s}&=& 
\frac{\ri}{2}  H_{\ga_1 \dots \ga_{2s}} D^2 H^{\ga_1 \dots \ga_{2s}} 
+ \frac{\ri}{2}  D^\b H_{\b \ga_1\dots  \ga_{2s-1}} D_\g H^{\g \ga_1\dots \ga_{2s-1}} 
\non  \\ 
\nonumber
&&+(2s-1) Y_{\ga_1\dots \ga_{2s-2}  } \pl_{\b \g} 
H^{ \b \g \ga_1\dots \ga_{2s-2} } 
+ \frac{\ri}{2}  (2s-1) Y_{\ga_1 \dots \ga_{2s-2}} D^2 Y^{\ga_1 \dots \ga_{2s-2}}  \\ 
&&- {\ri} (s-1) (2s-1)  D^\b Y_{ \b \ga_1 \dots \ga_{2s-3} } 
 D_\g Y^{ \g \ga_1 \dots \ga_{2s-3}} ~.
 \label{L1-3}
\eea
The gauge transformation laws \p{GT1-1} turn into
\begin{subequations}\label{GT1-2L}
\bea 
\delta H_{\ga_1 \dots \ga_{2s}} &=& 2s D_{(\ga_1} \z_{\ga_2\dots \ga_{2s})}~, 
\label{GT1-2L-a}\\ 
\delta Y_{\ga_1 \dots \ga_{2s-2}} &=&
- D^\b \z_{\b \ga_1 \dots \ga_{2s-2}} ~, \label{GT1-2L-b} \\ 
\delta \F_{\ga_1 \dots \ga_{2s-3}} &=& \frac{1}{2}
\partial^{\b \g } \z_{\b \g  \ga_1\dots \ga_{2s-3}}~. \label{GT1-2L-c}
\eea
\end{subequations}

Upon inspecting \eqref{L1-3} one may see 
that the Lagrangian 
is also well defined for the cases
$s=0$ and $s=1$ which have been excluded from the above consideration. 
For $s=0$ only the first term in the right-hand side of \eqref{L1-3} remains, 
and the resulting Lagrangian corresponds to 
the massless scalar multiplet described by the action \eqref{5.6}.
In the $s=1$ case, \eqref{L1-3} coincides with the gravitino multiplet Lagrangian 
\eqref{LFL}.

Since the Lagrangian \eqref{L1-3} defines an off-shell massless supermultiplet
for every integer $s=0, 1, 2, \dots$, we may introduce a 
generating formulation for the massless multiplets of arbitrary superspin. 
It is described by the Lagrangian 
\bea 
{\cal L}&=& \frac{\ri}{2} \langle H | (-1)^{N/2}(D^2 - N^{-1} \gamma^+ \g   ) | H \rangle 
\non \\
&&+\frac{\ri}{2} \langle Y |  (-1)^{N/2}\g^2   N^{-1} |H \rangle
- \frac{\ri}{2} \langle H | N^{-1} \g^{+2}   (-1)^{N/2}| Y \rangle  \non \\
&&-\frac{\ri}{2}\langle Y |  (-1)^{N/2}( D^2 +\g^+ \g) (N +2)^{-1}| Y \rangle \non  \\
&&= \sum_{s=0}^\infty \frac{1}{(2s)!}{\cal L}_{s}~,
\label{616}
\eea
in which the dynamical variables are given by 
\begin{subequations}
\bea
&&| H \rangle = \sum_{s=0}^{\infty} \frac{1}{(2s)!} H_{\ga_1 \dots \ga_{2s}} a^{\ga_1+ }  
\dots a^{\ga_{2s}+ } | 0\rangle~, 
\\ 
&&| Y \rangle = \sum_{s=0}^{\infty}
\frac{1}{(2s)!} {\cal G}_{\ga_1 \dots \ga_{2s}} a^{\ga_1+ }  
\dots a^{\ga_{2s}+ } | 0\rangle~, 
\eea
\end{subequations}
The action associated with \eqref{616} is invariant under 
gauge transformations of the form
\bea  
\delta | H \rangle =  \gamma^+ | \z \rangle ~, \qquad
 \delta | Y \rangle =  \gamma | \z \rangle ~, 
 \eea
 where the gauge parameter is
 \bea
 |\z  \rangle = \sum_{s=0}^{\infty}
 \frac{1}{(2s+1)!} \z_{\ga_1 \dots \ga_{2s+1}} a^{\ga_1+ }  \dots a^{\ga_{2s+1}+ } 
 | 0\rangle ~.
 \eea


\subsection{Half-integer superspin multiplets} \label{Is-l}

A Lagrangian formulation for a massless multiplet of 
half-integer superspin $(s+\hf)$, with $s>1$, 
contains a gauge superfield $| H_{2s+1} \rangle$, 
a compensator $| X_{2s-2} \rangle$ 
and an auxiliary superfield
$|\J_{2s-3} \rangle$.
The superfield Lagrangian, ${\cal L}^{\parallel}_{s+\hf}$, is 
\bea
 \nonumber 
\frac{(-1)^s}{(2s)!} {\cal L}^{\parallel}_{s+\hf}
&=&  \frac{\ri}{4}\langle H_{2s+1} | 2N \Box +{\ri} K_1 D^2
- \g^{+2} \g^2
| H_{2s+1} \rangle 
\\ 
&& +\frac{\ri}{2} \langle H_{2s+1} |  \gamma^{+3} | X_{2s-2} \rangle 
+\frac{\ri}{2}  \langle X_{2s-2} |  \gamma^3| H_{2s+1} \rangle 
\non \\
&& + \ri \langle X_{2s-2} |(N+2) D^2 | X_{2s-2} \rangle 
+\langle \J_{2s-3} | \gamma | X_{2s-2} \rangle 
\non \\
&&
+  \langle X_{2s-2} | \gamma^+ | \J_{2s-3} \rangle
- \ri \langle \J_{2s-3} | | \J_{2s-3} \rangle
~.
\label{L2-1}
\eea
The corresponding action proves to be invariant under the gauge transformations 
\begin{subequations} \label{GT2-1}
\bea
 \delta | H_{2s+1} \rangle &=& \ri \gamma^+ |\z_{2s} \rangle ~,\\ 
 \delta | X_{2s-2} \rangle &=& \frac{\ri}{2} \g^2   |\z_{2s} \rangle ~,\\ 
 \delta | \J_{2s-3} \rangle &=& \hf \gamma^3 |\z_{2s} \rangle ~.
\eea
\end{subequations}
The field $ | \J_{2s-3} \rangle$ can be integrated out using its equation of motion
\be
| \J_{2s-3} \rangle= -\ri \gamma | X_{2s-2} \rangle~.
\ee
Then the Lagrangian  \p{L2-1} turns into 
\bea \nonumber 
\frac{(-1)^s}{(2s)!}{\cal L}_{s+\hf}
&=&  \frac{\ri}{4}\langle H_{2s+1} | 2N \Box +{\ri} K_1 D^2
- \g^{+2} \g^2
| H_{2s+1} \rangle 
+\frac{\ri}{2} \langle H_{2s+1} |  \gamma^{+3} | X_{2s-2} \rangle 
\\ 
&& 
+\frac{\ri}{2}  \langle X_{2s-2} |  \gamma^3| H_{2s+1} \rangle 
- \ri \langle X_{2s-2} |\g^+ \g- (N+2) D^2 | X_{2s-2} \rangle ~.
 \label{L2-2}
\eea

Introducing the expansions
\begin{subequations}
\bea
|H_{2s+1} \rangle &=& \frac{1}{(2s+1)!} H_{\ga_1 \dots \ga_{2s+1}} a^{\ga_1+ }  
\dots a^{\ga_{2s+1}+ } | 0\rangle~, 
\\ 
|X_{2s-2} \rangle &=& \frac{1}{(2s-2)!} X_{\ga_1 \dots \ga_{2s-2}} a^{\ga_1+ }  
\dots a^{\ga_{2s-2}+ } | 0\rangle~, \\ 
|\J_{2s-3} \rangle &=& \frac{1}{(2s-3)!} \J_{\ga_1 \dots \ga_{2a-3}} a^{\ga_1+ }  \dots a^{\ga_{2a-3}+ } | 0\rangle
\eea
\end{subequations}
for the fields, and 
\be
|\z_{2s}  \rangle = \frac{1}{(2s)!} \z_{\ga_1 \dots \ga_{2s}} a^{\ga_1+ }  \dots a^{\ga_{2s}+ } | 0\rangle 
\ee
for the gauge parameters,
one gets from \p{L2-1} the Lagrangian in terms of the three fields
\bea \nonumber
(-1)^s 
{\cal L}^{\parallel}_{s+\hf}
&=& 
\frac{\ri}{2}H_{\ga_1\dots  \ga_{2s+1}} \Box H^{\ga_1 \dots  \ga_{2s+1}} 
- \frac{1}{4} H_{\ga_1\dots \ga_{2s} \b} \pl^{\b}{}_{ \g} D^2 H^{ \g \ga_1\dots \ga_{2s} }  
 \\ \nonumber
&&- \frac{\ri}{2} s\pa^{\b\g} H_{\b\g \ga_1\dots  \ga_{2s-1} }  
\pl_{\d \l} H^{\d\l \ga_1\dots  \ga_{2s-1} } 
 \\ \nonumber
&&-2s (2s-1) X_{\ga_1 \dots  \ga_{2s-2}}
\pl_{\b\g} D_\d H^{\b\g \d \ga_1 \dots  \ga_{2s-2} }   \\ 
&&+ \ri (2s)^2(2s-1) X_{\ga_1 \dots \ga_{2s-2}} D^2  X^{\ga_1 \dots  \ga_{2s-2}}  \non\\ \nonumber
&& -  \ri 2s (2s-1) (2s-2)\J_{\ga_1 \dots  \ga_{2s-3}} 
\J^{\ga_1 \dots \ga_{2s-3}}  \\ 
&&+4s (2s-1) (2s-2) \J_{\ga_1 \dots \ga_{2s-3}}
D_{\b} X^{\b \ga_1 \dots \ga_{2s-3}}
 \label{L2-3}
\eea
from \p{L2-2} the Lagrangian in terms of two fields
\bea \nonumber 
(-1)^s {\cal L}_{s+\hf}
&=& \frac{\ri}{2}H_{\ga_1\dots \ga_{2s+1}} \Box
H^{\ga_1\dots \ga_{2s+1}} 
- \frac{1}{4} H_{\ga_1\dots \ga_{2s} \b} \pl^{\b}{}_{ \g} D^2 H^{\g \ga_1\dots \ga_{2s} }  
 \\
  \nonumber
&& -\frac{\ri}{2} s \pa^{\b\g} H_{\b\g \ga_1\dots \ga_{2s-1}}  
\pl_{\d \l} H^{\d \l \ga_1\dots  \ga_{2s-1} } \\ 
&&-2s (2s-1) X_{\ga_1 \dots  \ga_{2s-2}}
\pl_{\b\g} D_\d H^{\b\g \d \ga_1 \dots  \ga_{2s-2} }  \non  \\ 
&&+ \ri (2s)^2(2s-1) X_{\ga_1 \dots \ga_{2s-2}} D^2  X^{\ga_1 \dots  \ga_{2s-2}}  \non\\ 
&&+ 4 \ri s (2s-1)(2s-2) D^\b X_{\b \ga_1\dots  \ga_{2s-3}}  D_\g  X^{\g \ga_1 \dots  \ga_{2s-3}}~.
\label{L2-3-1}
\eea
From \p{GT2-1} we read off the gauge transformations
\begin{subequations} \label{GT1-2}
\bea
\delta H_{\ga_1 \dots \ga_{2s+1} } &=& 
\ri(2s+1)  D_{(\ga_1} \z_{\ga_2 \dots \ga_{2s+1})} ~, \label{GT1-2a}\\ 
\delta X_{\ga_1 \dots \ga_{2s-2}} &=& -\frac{1}{2}\pl^{\b\g } 
\z_{\b \g \ga_1 \dots \ga_{2s-2}}~, \label{GT1-2b}\\ 
\delta \J_{\ga_1 \dots \ga_{2s-3}} &=& \frac{\ri}{2}\pl^{\b\g} D^\d 
\z_{\b \g\d \ga_1 \dots \ga_{2s-3}}~.
\eea
\end{subequations}

It can be seen that the Lagrangian \eqref{L2-3-1} is well defined for
the cases $s=0$ and $s=1$ excluded from the above consideration. 
For $s=0$ it coincides (modulo an overall factor of) with the Lagrangian 
for the vector multiplet, eq. \eqref{5.7}. For $s=1$ it coincides 
(modulo an overall factor) with the Lagrangian 
for the supergravity multiplet, eq. \eqref{LHL}.

Generating formulation 
\bea \nonumber 
{\cal L}
&=&  \frac{\ri}{4}\langle H | (-1)^{(N-1)/2}(2N \Box +{\ri} K_1 D^2
- \g^{+2} \g^2 )N^{-1}| H \rangle \non \\
&&+\frac{\ri}{2} \langle H | (-1)^{(N-1)/2} N^{-1}\gamma^{+3} | X \rangle 
+\frac{\ri}{2}  \langle X |  \gamma^3 (-1)^{(N-1)/2} N^{-1}| H \rangle \non \\
&& + \ri \langle X |(-1)^{N/2}(\g^+ \g- (N+2) D^2 ) (N +3)^{-1}| X \rangle  \non \\
&=&\sum_{s=0}^\infty\frac{1}{(2s+1)!} {\cal L}_{s+\hf}
~,
\eea
where 
\begin{subequations}
\bea
&&| H \rangle = \sum_{s=0}^{\infty} \frac{1}{(2s+1)!} H_{\ga_1 \dots \ga_{2s+1}} a^{\ga_1+ }  
\dots a^{\ga_{2s+1}+ } | 0\rangle~, 
\\ 
&&|X \rangle = \sum_{s=0}^{\infty}
\frac{1}{(2s)!} g_{\ga_1 \dots \ga_{2s}} a^{\ga_1+ }  
\dots a^{\ga_{2s}+ } | 0\rangle~, 
\eea
\end{subequations}
Gauge transformation 
\bea  
\delta | H \rangle = \ri  \gamma^+ | \z \rangle ~, \qquad
 \delta | X \rangle = \frac{\ri}{2} \gamma^2 | \z \rangle ~, 
 \eea
 where the gauge parameter is
 \bea
 |\z  \rangle = \sum_{s=0}^{\infty}
 \frac{1}{(2s)!} \z_{\ga_1 \dots \ga_{2s}} a^{\ga_1+ }  \dots a^{\ga_{2s}+ } 
 | 0\rangle ~.
 \eea

\subsection{Transverse formulation}

In this subsection, we briefly describe  massless gauge theories
realised in terms of the dynamical variables 
$\cV^{\perp}_{s}$ and  $\cV^{\perp}_{s+\hf}$ 
defined by eqs. \eqref{4.233} and \eqref{4.211}, respectively.

\subsubsection{Integer superspins}

A Lagrangian formulation for a massless multiplet of integer superspin $s$, with $s>1$, 
contains a gauge superfield $| H_{2s} \rangle$, a compensator $| Y_{2s-2} \rangle$ 
and an auxiliary superfield $| {\Phi}_{2s-1} \rangle$. 
The Lagrangian for this supermultiplet, ${\cal L}^\perp_s$,
is 
\bea \label{L1-1T} 
\frac{(-1)^s}{(2s-1)!} {\cal L}^\perp_s&=& \frac{\ri}{2} \langle H_{2s} | ND^2 | H_{2s} \rangle  - 
\ri \langle Y_{2s-2} | \gamma^2 | H_{2s} \rangle  
+\ri \langle H_{2s} | \gamma^{+2}| Y_{2s-2}  \rangle \\ \nonumber
&&+ \langle H_{2s}| \gamma^+ | \Phi_{2s-1} \rangle
+ \langle \Phi_{2s-1} | \gamma | | H_{2s} \rangle 
+\frac{\ri }{2}\langle Y_{2s-2} | (N+2)D^2 | Y_{2s-2} \rangle \\ \nonumber
&&-2\ri \langle \Phi_{2s-1} || \Phi_{2s-1} \rangle 
 - 
\langle Y_{2s-2}|\gamma | \Phi_{2s-1} \rangle + \langle \Phi_{2s-1}  |\gamma^+ | Y_{2s-2}\rangle~.
\eea
The corresponding action is  invariant under gauge transformations
\begin{subequations} \label{GT1-1T}
\bea
 \delta | H_{2s} \rangle &=& \gamma^+ | \z_{2s-1} \rangle ~,\\ 
 \delta | Y_{2s-2} \rangle &=& \gamma | \z_{2s-1} \rangle ~,\\ 
 \delta | {\Phi }_{2s-1} \rangle &=& (K_1 + \frac{\ri}{2}D^2)| \z_{2s-1}\rangle ~.
\eea
\end{subequations}
The auxiliary superfield  $| {\Phi }_{2s-1} \rangle$  cane be integrated out using 
its equation of motion
\be
| {\Phi }_{2s-1} \rangle = -\frac{\ri}{2} (\gamma | H_{2s} \rangle + \gamma^+ | Y_{2s-2} \rangle)~.
\ee
Then the Lagrangian \p{L1-1T} reduces to \p{L1-2}.

The fields and the gauge parameters can be expanded in terms of the oscillators, 
in complete analogy with our analysis in  subsection \ref{His-l}.
Then the gauge transformation laws \p{GT1-1T} turn into 
\begin{subequations}\label{GT1-2T}
\bea 
\delta H_{\ga_1 \dots \ga_{2s}} &=& 2s D_{(\ga_1} \z_{\ga_2 \dots \ga_{2s})} ~,\\ 
\delta {Y}_{\ga_1  \dots \ga_{2s-2}} &=& -D^\b \z_{\b \ga_1  \dots \ga_{2s-2}}~, \\ 
\delta {\Phi}_{\ga_1 \dots \ga_{2s-1}} &=& - 2s\partial^\b{}_{(\ga_1}
\z_{ \ga_2 \dots \ga_{2s-1}) \b} 
+\frac{\ri}{2} D^2  \z_{\ga_1 \dots \ga_{2s-1}}~,
\eea
\end{subequations}
and  the Lagrangian \p{L1-1T} becomes
\bea
(-1)^s{\cal L}^\perp_s&=& \frac{\ri}{2} H_{\ga_1\dots \ga_{2s}} D^2 
H^{\ga_1\dots \ga_{2s}} 
+ 2  (2s-1)Y_{\ga_1 \dots \ga_{2s-2}} \partial_{\b \g} 
H^{\b\g \ga_1 \dots \ga_{2s-2}} \non  \\ \nonumber
&&+2  \Phi_{\ga_1 \dots \ga_{2s-1}} D_\b  H^{\b \ga_1 \dots \ga_{2s-1}}
+{\ri} s  (2s-1) Y_{\ga_1 \dots \ga_{2s-2}} D^2 Y^{\ga_1 \dots \ga_{2s-2}}   \\ 
&& -2{\ri} \Phi_{\ga_1 \dots \ga_{2s-1}}  \Phi^{ \ga_1 \dots  \ga_{2s-1}}  
-2  (2s-1) Y _{\ga_1 \dots \ga_{2s-2}} D_\b \Phi^{ \b \ga_1 \dots \ga_{2s-2}} ~.
 \label{L1-3-0T}
\eea

\subsubsection{ Half-integer superspins}

A Lagrangian formulation for a massless multiplet of 
half-integer superspin $(s+\hf)$, with $s>1$, 
contains a gauge superfield $| H_{2s+1} \rangle$, 
a compensator $| X_{2s-2} \rangle$ 
and an auxiliary superfield
$|\J_{2s-1} \rangle$.
The  Lagrangian, ${\cal L}^\perp_{s+\frac{1}{2}}$, is
\bea  \label{L2-1T}
\frac{(-1)^s}{(2s)!}{\cal L}^\perp_{s+\frac{1}{2}} &=& \frac{\ri}{4}\langle H_{2s+1}| 2N \Box  +i K_1 D^2   | H_{2s+1}  \rangle   
 +  \langle H_{2s+1}  | \gamma^{+2}| \Psi_{2s-1} \rangle  \\ \nonumber
&-& \langle \Psi_{2s-1}| \gamma^2 | H_{2s+1} \rangle  
+ \langle H_{2s+1} | \gamma^{+3} | X_{2s-2} \rangle +\langle X_{2s-2}|  \gamma^3| H \rangle  \\ \nonumber
&+& \ri \langle X_{2s-2} | D^2 | X_{2s-2} \rangle 
+4 \ri \langle \Psi_{2s-1}| | \Psi_{2s-1} \rangle
- 2 \langle \Psi_{2s-1} | \gamma^+ | X_{2s-2} \rangle \\ \nonumber
&+&2 \langle X_{2s-2} | \gamma | \Psi_{2s-1} \rangle~.
\eea
The corresponding action is invariant under gauge transformations 
\begin{subequations} \label{GT2-1T}
\bea
 \delta | H_{2s+1} \rangle &=& \ri \gamma^+ |\z_{2s} \rangle ~,\\ 
 \delta | X_{2s-2}\rangle &=& \frac{\ri}{2} \gamma^2 | \z_{2s} \rangle ~,\\ 
\delta | \Psi_{2s-1} \rangle &=&  
\left ( \frac{1}{2}\gamma^+ \gamma^2  +\frac{\ri}{2} P \right ) |\z_{2s} \rangle ~.
\eea
\end{subequations}
 The auxiliary superfield
$ | \Psi _{2s-1}\rangle$  
can be integrated out using it equation of motion
\be
| \Psi_{2s-1} \rangle=- \frac{\ri}{4} \gamma^2 | H_{2s} \rangle - \frac{\ri }{2}\gamma^+| X_{2s-2}\rangle    ~,
\ee
which turns
the Lagrangian  \p{L2-1T} into  \p{L2-2}.

The above results can be rewritten in terms of ordinary superfields
(compare with subsection \ref{Is-l}).
The gauge transformation laws \p{GT2-1T} are equivalent to 
\begin{subequations}\label{GT1-2T-1}
\bea 
\delta H_{\ga_1  \dots \ga_{2s+1} } &=& 
(2s+1)\ri  D_{(\ga_1} \z_{\ga_2 \dots \ga_{2s+1})}~, \\ 
\delta X_{\ga_1  \dots\ga_{2s-2}} &=& 
- \frac{1}{2}\pl^{\b \g} \z_{\b \g \ga_1  \dots \ga_{2s-2}} ~,\\ 
\delta \Psi_{\ga_1 \dots \ga_{2s-1}} &=& \frac{\ri}{2} (2s-1)
\pa^{\b\g} D_{(\ga_1}   \z_{ \ga_2 \dots \ga_{2s-1}) \b\g} + \frac{\ri}{2} 
\partial^{\b\g}D_\b \z_{\g \ga_1 \dots \ga_{2s-1}}~.
\eea
\end{subequations}
The Lagrangian \p{L2-1T} is equivalent to 
\bea 
(-1)^s{\cal L}_{s+\frac{1}{2}}&=& \frac{\ri}{2}H_{\ga_1 \dots \ga_{2s+1}} 
\Box H^{\ga_1\dots \ga_{2s+1}} 
- \frac{1}{4} H_{\ga_1 \dots \ga_{2s} \b} \pl^{\b}{}_{ \g} D^2 H^{\g \ga_1 \dots \ga_{2s} }   \non \\ 
&& -4 \ri s\Psi_{\ga_1 \dots \ga_{2s-1}}\pl_{\b \g }H^{\b \g \ga_1 \dots \ga_{2s-1}}   
-4s (2s-1)X_{\ga_1 \dots\ga_{2s-2}} \pl_{\b \g }D_{\d} H^{\b\g\d \ga_1 \dots\ga_{2s-2}} 
\non \\ 
&& +2 \ri s (2s-1)X_{\ga_1\dots\ga_{2s-2}}D^2 X^{\ga_1\dots\ga_{2s-2}}  
+8 \ri s \Psi_{\ga_1 \dots \ga_{2s-1}}\Psi^{  \ga_1 \dots \ga_{2s-1}}  \non \\ 
&& +8s(2s-1)X_{\ga_1 \dots\ga_{2s-2}} D_\b \Psi^{ \b \ga_1 \dots\ga_{2s-2}}~.
 \label{L2-3-4}
\eea


\section{Massive higher spin supermultiplets}

Before presenting the Lagrangians for massive supermultiplets
and analysing the corresponding equations of motion,
it is useful to reformulate some of the results given
in subsection 3.2 in terms of the auxiliary oscillators used in the previous section.

\subsection{Higher spin super-Cotton tensor}\label{subsection7.1}

One can check that the 
higher spin super-Cotton tensor
\p{2.32} 
can be written in the form
\be \label{GC-1}
| W_n \rangle = (-1)^n \left ( \sum_{p=0}^{[n/2]} a_p  \Box^p K_{n-2p}
+ \ri \sum_{p=0}^{[n/2]} b_p  \Box^pD^2 K_{n-2p-1}   \right ) | H_n \rangle~.
\ee
The expression \p{GC-1} is invariant under gauge transformations
\be \label{TTT}
\delta |H_n \rangle = \gamma^+ | \Lambda_{n-1} \rangle
\ee
provided the constant coefficients $a_p$ and $b_p$ satisfy the equations
\bea \label{abc}
&&a_p (n-2p) - 2  b_p=0~, \qquad a_{p+1} - 2   b_p (n-2p-1)=0~.
\eea
These recurrence relations are solved by 
\be \label{solab}
a_p =  \binom{n}{2p} (2p)!a_0~,  \qquad  b_p = \frac{ 1 }{2} \binom{n}{2p+1}
(2p+1)! a_0~.
\ee
In order to match the  overall coefficient in \p{2.32}, $a_0$ has to be 
\bea
a_0 = \frac{1}{n! 2^n} ~.
\eea
One may check that the gauge-invariant field strength  \p{GC-1}  
satisfies the Bianchi identity
\bea \label{BBB}
\gamma | W_n \rangle=0~.
\eea

\subsection{Superfield Lagrangian}

Now let us turn to describing our off-shell massive higher spin $\cN=1$ supermultiplets.
The Lagrangian for a massive superspin-$\frac{n}{2}$ multiplet
 is defined by
\be \label{LM-1}
{\cal L}^{(n/2)}_{\rm massive} = \frac{1}{(n-1)!}{\cal L}_{n/2} + \frac{\ri^n }{2}
n\lambda ( \langle H_n || W_n \rangle + \langle W_n || H_n \rangle )~,
\ee
where the massless Lagrangian ${\cal L}_{n/2} $ is given either by the equation 
 \p{L1-2} for $n=2s$,
  or by  \p{L2-2} for $n=2s+1$.
  (We recall that the oscillator realisation for the super-Cotton tensor 
 $| W_n \rangle$
is described  
in subsection \ref{subsection7.1}.)
Thus the massive Lagrangian 
is obtained from the massless one
by adding the Chern-Simons like term.

The action generated by \p{LM-1} is gauge invariant since both terms 
in the action are separately gauge invariant.
Indeed the term proportional to $\lambda$ is invariant under the transformations \p{TTT}
due to the gauge invariance of $| W_n \rangle$ and the Bianchi identities \p{BBB}.
The gauge invariance of 
the first term was established in section \ref{section6}.
Note also that the mass term contains only the physical gauge superfield $| H_n \rangle$
and does not depend on the compensator.

In the integer superspin case, $n=2s$, the gauge-invariant equations of motion 
 derived from \p{LM-1} are
\begin{subequations}\label{EOMM-1-1-2-1}
\bea \label{EOMM-1-1}
| E_{2s} \rangle
 + 2s \lambda | W_{2s} \rangle  &=&0~,\\
\label{EOMM-2-1}
| Q_{2s-2} \rangle&=&0~, 
\eea
\end{subequations}
where we have introduced the following gauge-invariant 
  field strengths:
\begin{subequations}
\bea \label{Eodd}
| E_{2s} \rangle &=& \frac{\ri}{2} (N D^2 - \gamma^+ \gamma )   | H_{2s} \rangle 
+ \frac{\ri}{2} \gamma^{+ 2}
 | {Y}_{2s-2}\rangle  ~,\\
\label{Qodd}
| Q_{2s-2} \rangle &=& -\frac{\ri}{2}\gamma^2  | H_{2s} \rangle + \frac{\ri}{2}(D^2 + \gamma^+ \gamma)| {Y}_{2s-2}\rangle~.
\eea
\end{subequations}
These field strengths are gauge invariant, since 
they are proportional to the equations of motion
for the massless model $\cL_s$. 
The field strengths $| E_{2s} \rangle $ and $| Q_{2s-2} \rangle $ 
obey the Bianchi identity
\be \label{B1}
\gamma | E_{2s} \rangle  = \gamma^+ | Q_{2s-2} \rangle~,
\ee
which expresses the gauge invariance of the massless action.
Therefore the field strength $ | E_{2s}\rangle$ is transverse
provided the equation of motion  \p{EOMM-2-1} holds, that is
\be \label{gammaE}
| Q_{2s-2} \rangle=0 \quad \Longrightarrow \quad
\gamma | E_{2s}\rangle =0~.
\ee
We point out that 
 \p{EOMM-2-1}
 is the equation of motion for 
 the compensator $| Y_{2s-2} \rangle $.

One can consider the half-integer superspin case in a similar way.  
The corresponding equations of motion are
\begin{subequations}\label{EOMM-3-4}
\bea 
| E_{2s+1} \rangle
   + (-1)^s 
  (2s+1){ \lambda} | W_{2s+1} \rangle  &=&0~,\label{EOMM-3}  \\
\label{EOMM-4}
| R_{2s-2} \rangle 
&=&0~,
\eea
\end{subequations}
where we have introduced the gauge-invariant field strengths:
\begin{subequations}
\bea 
\label{Even}
| E_{2s+1} \rangle &=& \frac{1}{4} (2  N \Box + \ri K_1D^2 - \gamma^{+2} \gamma^2)   | H_{2s+1} \rangle +\frac {1}{2} \gamma^{+3}| X_{2s-2} \rangle ~,\\
\label{Qeven}
| R_{2s-2} \rangle &=& -\frac{\ri}{2}\gamma^3  | H_{2s+1}\rangle 
-  \ri((N+2)D^2 - \gamma^+ \gamma)| {X}_{2s-2}\rangle~ .
\eea
\end{subequations}
The Bianchi identity relating the  field strengths  \p{Even} and  \p{Qeven} reads
\be \label{B2}
\gamma | E_{2s+1} \rangle  = -\frac {\ri}{2} \gamma^{+2} | R_{2s-2} \rangle~.
\ee
Therefore the field strength $| E_{2s+1} \rangle$ satisfies 
 the equation \p{gammaE} provided $| R_{2s-2} \rangle =0$.

Now we are in a position to analyse equations of motions for massive supermultiplets.
We start with gravitino and supergravity multiplets and then generalise the analysis for
an arbitrary superspin.


\subsection{Massive gravitino multiplet}
For massive gravitino multiplet one has the following expressions for the fields strength
$| E_2 \rangle$ and $| Q_0\rangle$:
\begin{subequations}
\bea 
\label{Egravitino}
| E_2 \rangle &=& \frac{\ri}{2} (D^2 - \ri K_1)   | H_2 \rangle + \frac{\ri}{2} \gamma^{+2} | {Y}_0\rangle ~, \\
 \label{Qgravitino}
| Q_0 \rangle & = & -\frac{\ri}{2} \gamma^2| H_2 \rangle 
+ \frac{\ri}{2} D^2| {Y}_0\rangle ~.
\eea
\end{subequations}
Taking the linear combination
\be
|W_2 \rangle:=
\frac {\ri}{4} (D^2 | E_2 \rangle -\gamma^{+2} | Q_0\rangle )
\ee
and using the expressions \p{ll} and \p{k1kl}
one obtains
\be
|W_2 \rangle=\frac{1}{8} (K_2 +2 \Box + \ri K_1 D^2) | H_2 \rangle~,
\ee
which is the same as the superconformal field strength  \p{GC-1}
for $n=2$.

Let us analyse  the  equations of motion. As mentioned above,
 the equation of motion \p{EOMM-2-1} for
the compensator is $| Q_0 \rangle =0$. Furthermore,
the equation of motion for the field strength  $| E_2 \rangle$ is
\be \label{E2}
|E_2 \rangle +\frac{\ri}{2} \lambda D^2 | E_2 \rangle =0
\ee
and, due to the Bianchi identity \p{B1}, 
the field strength satisfies $\gamma | E_2 \rangle =0$.
The equation \p{E2} in turn implies
\be
(\Box - m^2) | E_2 \rangle =0~, \qquad m^2=\frac{1}{ \lambda^2}~.
\ee

It is important to notice that since the Chern-Simons like mass term in \p{LM-1}
does not contain compensator superfields one can immediately recover two dually invariant formulations
for the gravitino multiplet, alike those in the subsection \ref{section5.2}.
Explicitly the field strength has the form
\be \label{CURVB-2}
W_{\alpha_1 \alpha_2}= -\frac{1}{4} D^{\beta_1} D_{\alpha_1} D^{\beta_2} D_{\alpha_2} H_{\gb_1 \gb_2}
\ee
and, therefore, the Lagrangian for the massive  gravitino multiplet has the form
\label{LFL-m}
\bea
L_{\rm GM} &=& -\frac{\ri}{2} \Big\{  H^{\alpha \beta} D^2 H_{\alpha \beta}
+ D_\alpha H^{\alpha \beta} D_\gamma H^{\gamma}{}_{\beta}
- 2  D_{\a} H^{\alpha \beta} D_\b X
- D^\b X D_\b X \Big\}  \\ \nonumber
&&-{\lambda} H^{\alpha \beta}  W_{\alpha \beta} ~,
\eea
while the dual Lagrangian is 
\bea
L^{(dual)}_{\rm GM} &=&- \frac{\ri}{2} \Big\{   H^{\alpha \beta} D^2 H_{\alpha \beta} 
+ 2D_\alpha H^{\alpha \beta} D_\gamma H^{\gamma}{}_{\beta} 
+ 2\ri W^\a D^\b H_{\a\b}  - W^\a W_\a \Big\} \\ \nonumber
 &&-{\lambda} H^{\alpha \beta}  W_{\alpha \beta} 
~.~~
\eea

\subsection{Massive supergravity multiplet}
One can perform a similar procedure for the supergravity multiplet.
The corresponding field strengths are
\begin{subequations}
\bea
 \label{Egraviton}
| E_3 \rangle &=& \frac{1}{4}( 6  \Box + \ri K_1D^2 - \gamma^{+ 2} \gamma^2)   | H_3 \rangle 
+ \frac{1}{2} \gamma^{+3}| X_0 \rangle ~,\\
\label{Qgraviton}
| R_0 \rangle &=& -\frac{\ri}{2}\gamma^3 | H_3 \rangle - 2 \ri D^2 | {X}_0\rangle~.
\eea
\end{subequations}
Taking a linear combination
\be
| W_3 \rangle := \frac{1}{3!} ( -\ri D^2 | E_3 \rangle 
+   \gamma^{+3} | R_0 \rangle)~,
\ee
one obtains
\be
| W_3 \rangle = -\frac{1}{3! \cdot 8}(K_3 +6 K_1 \Box + \frac{3 \ri}{2}K_2 D^2 
+  3 \ri \Box D^2 )| H_3 \rangle~,
\ee
which is the field strength \p{GC-1} for $n=3$.
The equation of motion for the compensator is $| R_0 \rangle =0$, whereas
the equation of motion with respect to $| H_ 3\rangle$ is
\be
| E_3 \rangle + \frac{\ri}{2}\lambda  D^2 | E_3 \rangle =0
\ee
which in turn implies
\be
(\Box - m^2) | E_3 \rangle =0~, \qquad m^2=\frac{1}{ \lambda^2}~.
\ee

Similar to the case of the gravitino multiplet, one can 
present two dual formulations for the massive supergravity multiplet.
Since the linearised super-Cotton tensor is independent of the compensator,
\be \label{CURVH-1}
W_{\alpha_1 \alpha_2 \alpha_3}=\frac{\ri}{8} D^{\beta_1} D_{\alpha_1} D^{\beta_2} D_{\alpha_2} 
D^{\beta_3} D_{\alpha_3} H_{\gb_1 \gb_2 \gb_3}~,
\ee
one can write the Lagrangian for the massive supergravity multiplet
\bea \nonumber
L_{\rm SGM} &=& \frac{\ri}{4}  H^{\alpha \beta \gamma} \Box H_{\alpha \beta \gamma}
- \frac{1}{8}H^{\alpha \beta \gamma} \partial_{\gamma \rho} D^2
H_{\alpha \beta}{}^{ \rho}
- \frac{\ri}{4} 
\partial_{\a \b}H^{\a \b \gamma} \partial^{\rho \sigma}H_{\rho \sigma}{}_{\gamma} \\
&& +\hf \partial_{\a\b} H^{\a \b \gamma}D_\gamma X 
+ \frac{\ri}{2} D^\g X D_\g  X \\ \nonumber
&& +\ri\lambda H^{\alpha \beta \gamma} W_{\alpha \beta \gamma}~,
 \label{LHL-m}
\eea
as well as its dual form
\bea \nonumber 
L^{(dual)}_{\rm SGM}&=& \frac{\ri}{4}  H^{\alpha \beta \gamma} \Box H_{\alpha \beta \gamma}
- \frac{1}{8}H^{\alpha \beta \gamma} \partial_{\gamma \rho} D^2
H_{\alpha \beta}{}^{ \rho}
- \frac{\ri}{8} 
\partial_{\a \b}H^{\a \b \gamma} \partial^{\rho \sigma}H_{\rho \sigma}{}_{\gamma} \\
&& +\frac{\ri}{2} \partial_{\a\b} H^{\a \b \gamma}W_\g
+ \frac{\ri}{2} W^\a W_\a \\ \nonumber
&&+ \ri \lambda H^{\alpha \beta \gamma} W_{\alpha \beta \gamma}
~.
\eea


\subsection{Arbitrary superspin}

In order to analyse the case of an arbitrary integer superspin, let us consider 
 the gauge-invariant action for $n=2s$
\bea
S^{(s)}_{\rm massive}
=
 \int \rd^{3|2}z \,
\Big\{ 
\cL_{s} \big( H_{\a(2s)}, Y_{\a(2 s -2)}\big)
+(-1)^s \l H^{\a (2s) }  W_{\a (2s)} (H) \Big\}~,
\eea
where the Lagrangian $\cL_s$ is given by \eqref{L1-3}. Let us analyse the equations of motion. The gauge invariance \eqref{GT1-2L}
allows us to choose a gauge $Y_{\a(2s-2)}=0$ in which the equation of motion for 
$Y_{\a(2s-2)}$ amounts to $\pa^{\b\g} H_{\b\g \a(2s-2)} =0$, and the  residual gauge 
freedom is constrained by $D^\b \z_{\b \a(s2-2)} =0$. On the mass shell, 
the residual gauge freedom can 
be used to to impose a stronger condition on the gauge prepotential, 
\bea \label{stgc}
D^\b H_{\b \a(2s-1)}=0~.
\eea
In this gauge the super-Cotton tensor becomes
\bea
W_{\a(2s)} = \Box^s H_{\a(2s)}~,
\label{7.40}
\eea
in accordance with \eqref{3377}. Under the same gauge condition, the equation of motion 
for $H_{\a(2s)}$ reduces to
\bea
\frac{\ri}{2} D^2H_{\a(2s)}+ \l W_{\a(2s)} =0~.
\label{7.41}
\eea
Combining the equations \eqref{7.40} and \eqref{7.41} gives
\bea
\Box \Big(\Box^{2s-1} -\l^{-2} \Big) H_{\a(2s)}=0~.
\eea
If we choose the solution $\Box H_{\a(2s)}=0$, the super-Cotton tensor vanishes, 
$W_{\a(2s)}=0$, in accordance with \eqref{7.40}. Then the equations of motion reduce to 
the massless ones, which means the gauge field can be completely gauged away. 
Thus the nontrivial solutions obey the equations
\bea
\Big(\Box^{2s-1} -\l^{-2} \Big) H_{\a(2s)}=0~,
\label{7366}
\eea
which implies\footnote{Compare with a similar analysis in the $\cN=2$ case \cite{KO}.}
\bea \label{mass-sh}
\Big(\Box -m^2 \Big) H_{\a(2s)}=0~, \qquad 
m = \frac{1}{| \l | ^{1/(2s-1)}}~.
\eea
The  equations \eqref{7.40} and \eqref{7.41} also imply that $W_{\a(2s)}$
is an on-shell massive superfield in the sense of \eqref{214},
\bea
\Big( \frac{\ri}{2}  D^2 +m \s \Big) W_{\a(2s)} =0~, \qquad \s = (-1)^s \frac{\l}{|\l|}~,
\eea
and hence the superhelicity of $W_{\a(2s)}$ is 
$\k = \left(s +\frac{1}{4} \right) \s$.

It is instructive to repeat the above analysis by making use of the oscillator 
realisation.
The relation between 
the  field strength \p{GC-1} and $| E_{2s} \rangle$ is
\be \label{GC-1-1}
| W_{2s} \rangle =  \left (  \sum_{p=0}^{s-1} c_p  \Box^p K_{2s-2p-1}   \right ) | E_{2s} \rangle~,
\ee
where the coefficients $c_p$ are related to the coefficients $b_p$ given in  \p{solab} as
$c_p = \frac{1}{s}b_p$.
After gauging away the compensator $| {Y}_{2s-2} \rangle$,
  imposing the corresponding equation of motion $| Q_{2s-2} \rangle =0$ and
 the condition \p{stgc} on the gauge potential
\be\label{stgc-1}
\gamma | H_{2s} \rangle=0~,
\ee
one obtains
\be
|W_{2s} \rangle  = \Box^s | H_{2s} \rangle
\ee
and 
\bea \label{mass-sh-1}
\left ( \Box - \frac{2}{| \l | ^{1/(N-1)}}\right ) | H_{2s} \rangle=0~, 
\eea
with the latter being  the same  equation as 
\eqref{7366}.
The equation \p{mass-sh-1} implies that the field strength $| E_{2s} \rangle$  satisfies the same Klein-Gordon equation
\be \label{KG}
\left ( \Box - \frac{2}{| \l | ^{1/(N-1)}}\right )| E_{2s}\rangle =0~.
\ee

The analysis of the equations for the half-integer superspin case
simplifies due to an observation that 
there is a transformation that connects the systems of field equations 
for integer and half-integer superspins.
Again we are considering a partially gauge fixed system when the compensators $| {X}_{2s-2} \rangle$  and $| {Y}_{2s-2} \rangle$ are gauged away.
Then one can check
that the transformation 
\be \label{Darboux}
|H_{2s+1} \rangle = \xi P^+ |H_{2s} \rangle ~,
\ee
where $\xi$ is some Grassmann even constant parameter, transforms the solutions of the 
 equations \p{EOMM-1-1-2-1}
  into  solutions of the system \p{EOMM-3-4}
in the limit of zero mass (i.e., when $\lambda \rightarrow 0$). 
Moreover,
defining the operators  $E_{2s+1}$ and $E_{2s}$ as
\be
E_{2s+1} | H_{2s+1} \rangle =| E_{2s+1} \rangle, \quad E_{2s} | H_{2s} \rangle 
=| E_{2s}\rangle~,
\ee
one has the following chain of equations:
\be
E_{2s+1}  | H_{2s+1} \rangle = \xi
E_{2s+1} P^+  | H_{2s} \rangle =  -\frac{1}{4} \x P^+ \gamma^{+2} \gamma^2| H_{2s} \rangle =
 -\frac{\ri}{2} \xi P^+ \gamma^{+2} | Q_{2s-2} \rangle~.
\ee
Using the Bianchi identity \p{B1} one finally gets
\be \label{DM}
E_{2s+1}  | H_{2s+1} \rangle= | E_{2s+1} \rangle
=- \frac{\ri}{2} \xi P^+ \gamma^+ \gamma | E_{2s} \rangle ~.
\ee
After establishing this connection between field strengths for integer and half-integer superspins
one can multiply the equation
 \p{KG} with the operator $P^+ \gamma^+ \gamma$,
to obtain the result that  the field strength $|E_{2s+1} \rangle$ 
satisfies the Klein-Gordon equation
\be \label{KG-1}
\left ( \Box - \frac{2}{| \l | ^{1/(N-2)}}\right )| E_{2s+1}\rangle =0~,
\ee
as was the case of integer superspins.
From the equations \p{KG} and \p{KG-1}, one can see that
the fields with integer and half-- integer superspins have the same mass, as a result 
of the original $\cN=2$ supersymmetry.
Let us note however, that  as soon as one considers $\cN=1$ supersymmetry the parameter $\lambda$
does not have to be the same for integer and half-integer superspins.
Moreover for the case of free fields, which is our concern in the present paper, the parameter $\lambda$ can be different from each separate
value of a superspin, either it is  integer or  half--integer.


\section{Discussion}

In this paper we constructed the off-shell higher spin $\cN=1$ supermultiplets
in three dimensions, both in the massless and massive cases. 
Our massive actions are actually defined for arbitrary non-zero superspin.
They are labelled  
 by a positive integer, $n=1,2, \dots,$ and have the form 
\bea
S^{(n/2)}_{\rm massive}
=
 \int \rd^{3|2}z \,
\Big\{ 
\cL_{n/2} \big( H_{\a(n)}, \cX_{\a(2 {\left \lfloor{n/2}\right \rfloor} -2)}\big)
+\ri^n \l H^{\a_1 \dots \a_n }  W_{\a_1 \dots \a_n} (H) \Big\}~.
\label{8massive}
\eea
Here the compensator $ \cX_{\a(2 {\left \lfloor{n/2}\right \rfloor} -2)}$
is not present in the case $n=1$, which corresponds to 
the topologically massive vector multiplet. In section \ref{section6},
the compensator $ \cX_{\a(2 {\left \lfloor{n/2}\right \rfloor} -2)}$ 
was denoted $Y_{\a(2s-2)} $ for even $ n=2s$, and $X_{\a(2s+1)}$ 
for odd $n= 2s+1$.
The cases $n=2$ and $n=3$ correspond to the topologically massive gravitino and supergravity multiplets, respectively. The action \eqref{8massive} is gauge invariant.
It may be shown that the massless actions
\bea
S^{(n/2)}_{\rm massless}
=
 \int \rd^{3|2}z \,
\cL_{n/2} \big( H_{\a(n)}, \cX_{\a(2 {\left \lfloor{n/2}\right \rfloor} -2)}\big)
\eea
do not describe any propagating degrees of freedom for $n>1$. 
Nontrivial dynamics in the massive case is due to the presence of  
the Chern-Simons like term \eqref{8massive}.

In section \ref{section5}, we constructed two dual formulations 
for the massless gravitino multiplet and for the  linearised supergravity multiplet.
Deforming the dual massless actions by  Chern-Simons like mass terms according to  \eqref{8massive}, we end up with two dual formulations for 
the corresponding massive multiplets.
At the nonlinear level, only one off-shell formulation for $\cN=1$ supergravity 
has been constructed so far, and its conformal compensator is a scalar superfield, 
see \cite{KLT-M11} for a review. The fact that we now have two different off-shell actions
for linearised supergravity is intriguing, and it may imply the existence of 
a new off-shell supergravity formulation.

Our massive supermultiplets can be coupled to  external sources 
$\cJ_{\a(n)}$ using an action functional of the form
\bea
S^{(n)}_{\rm massive} 
\big[ H_{\a(n)}, \cX_{\a(2 {\left \lfloor{n/2}\right \rfloor} -2)}\big]
+  \ri^n \int \rd^{3|2}z \,
H^{\a_1 \dots \a_n } 
\cJ_{\a_1 \dots \a_n} 
~.
\eea
In order for such an  action to be invariant under the gauge transformations 
\eqref{GT1-2L-a} and \eqref{GT1-2L-b} for $n=2s$ or  under the gauge transformations 
\eqref{GT1-2a} and \eqref{GT1-2b} for $n=2s+1$, the source must be conserved, 
that is 
\bea
D^\b \cJ_{\b \a_1 \dots \a_{n-1} }=0~.
\label{8.2}
\eea
Such a superfield contains two ordinary conserved currents \cite{NSU},
which can be chosen as
\bea
j_{\a_1 \dots \a_n}(x) = \cJ_{\a_1 \dots \a_n}\Big|_{\q=0}~, \qquad 
j_{\a_1 \dots \a_{n+1} }(x) = \ri^{n+1} D_{(\a_1}\cJ_{\a_2 \dots \a_{n+1})}\Big|_{\q=0}~.
\eea
It follows from \eqref{8.2} that 
\bea
\pa^{\b\g} j_{\b\g \a_1\dots \a_{n-2}} =0~, \qquad \pa^{\b\g} j_{\b\g \a_1\dots \a_{n-1}} =0~.
\eea
In  3D $\cN=1$ superconformal field theory,  $\cJ_{\a\b\g}$ describes 
the supercurrent multiplet, $\cJ_\a$ is present if the theory 
possesses a flavour symmetry, 
and  $\cJ_{\a\b} $ emerges if the theory possesses an extended supersymmetry,
see \cite{BKS} for more details.

In the $\cN=2$ supersymmetric case, the massive higher spin supermultiplets 
constructed in \cite{KO} are gauge theories with linearly 
dependent generators, following the terminology of the Batalin-Vilkovisky 
quantisation \cite{BV} (see  \cite{Henneaux:1992ig} for a pedagogical review).
The Lagrangian quantisation of such gauge theories is nontrivial.\footnote{This is 
similar to the off-shell 4D $\cN=1$ massless higher spin supermultiplets \cite{KSP,KS93}, 
which are also reducible gauge theories.  
For the off-shell $\cN=1$ massless higher spin supermultiplets in AdS${}_4$ \cite{KS94}, 
the Lagrangian quantisation was carried out in \cite{BKS95}.}
The remarkable feature of our 3D $\cN=1$ massive higher spin supermultiplets
is that they are irreducible gauge theories that can be quantised using the 
standard Faddeev-Popov procedure.

Our construction of the massive higher spin supermultiplets may be viewed
as a generalisation 
of the topologically massive vector multiplet model \cite{Siegel}
\bea
S_{\rm TMVM} &=&- \frac{\ri}{2} \int \rd^{3|2} z \,  W^\a W_\a 
-\frac{\ri}{2} m \s  \int \rd^{3|2} z \,  H^\a W_\a ~,
\label{8.5}
\eea
where $\s=\pm 1$. The equation of motion in this theory is 
\bea
-\frac{\ri}{2} D^2  W_{\a} &=& m \s W_\a~. 
\eea
In conjunction with the Bianchi identity $D^\a W_\a=0$, this amounts
to \eqref{214} with $n=1$. Similar to \eqref{8.5}, our higher spin gauge theories
describe irreducible massive supermultiplets that propagate a single superhelicity state.
For low spin fields, however, there is a more traditional way to
generate off-shell massive supermultiplets that are parity-invariant and, therefore,
do not describe a single superhelicity. They are extensions 
of the massive vector multiplet model
\bea
S_{\rm MVM} &=&- \frac{\ri}{2} \int \rd^{3|2} z \,  (W^\a W_\a -m^2 H^\a H_\a)~,
\label{8.7}
\eea
in which the mass term involves the naked prepotential $H_\a$ squared such that
the action is not gauge invariant. The equation of motion for this action is
\bea
0= -\frac{\ri}{2}  D^\b D_\a W_\a +m^2 H_\a = \pa_\a{}^\b W_\b + m^2 H_\a~,
\eea
which implies
\bea
D^\a H_\a =0~, \qquad (\Box -m^2)H_\a =0~.
\eea
Due to the identity 
\bea
(\Box -m^2) = \Big(\frac{\ri}{2} D^2 +m \Big) \Big(\frac{\ri}{2} D^2 -m \Big)~,
\eea
it follows that the theory propagates two irreducible on-shell multiplets 
with superhelicity values $\k = \pm 3/4$, compare with \eqref{218}.
In the case of 4D $\cN=1$ Poincar\'e supersymmetry, 
there have appeared various off-shell realisations 
for the massive gravitino and supergravity multiplets
\cite{OS2,BGLP,GSS,BGKP,Gates:2006cq,Gates:2013tka}
that, conceptually, are similar to \eqref{8.7}. Analogous massive models 
without gauge invariance may be constructed
in the 3D $\cN=1$ case as well. An interesting point is that our
 {\it off-shell} massless 3D $\cN=1$ higher spin supermultiplets 
 appear to be  suitable to lift 
 the component {\it on-shell} massive gauge-invariant construction of \cite{BSZ} to superspace. 
 
A topic of our particular interest is an application of our results to the systems of interacting
higher spin fields on AdS backgrounds  first developed in 
\cite{Vasiliev:2003ev}--\cite{Prokushkin:1998bq},
which have received much interest in the last years. 
In relation to higher spin gauge theories
our results can be a step towards a few further developments which we hope do address in  future work. We conclude by listing possible future lines of work:

\begin{itemize}

\item
The massive higher spin supermultiplets constructed 
in this paper can be extended 
to 3D $\cN=1$ AdS superspace\footnote{To appear soon.} AdS${}^{3|2}$  (defined, e.g., in  \cite{KLT-M12}). 
It would be interesting to quantise the gauge-invariant massive  higher spin 
theories in AdS${}^{3|2}$  and to compute the corresponding partition functions.

\item It would be interesting to construct a BRST formulation for these systems both on
flat and AdS${}_3$ backgrounds. We would like to mention that in this respect, 
 in terms of their structure, the field equations for the integer and half-integer
3D ${\cal N}=1$
 higher spin supermultiplets
are very similar to  so--called triplet formulations for 
 massless \cite{Ouvry:1986dv}--\cite{Agugliaro:2016ngl} and massive \cite{Hussain:1988uk}--\cite{Bekaert:2003uc}
reducible higher spin fields, usually formulated in terms of the BRST formalism
(see also  \cite{Metsaev:2014vda}--\cite{Metsaev:2015oza} for a recent work on  
BRST-FV  approach for massive and massless higher spin fields).
Indeed, in both cases the Lagrangian system of equations contains a physical field and two auxiliary fields.
One of these fields is eliminated via its own equation of motion, whereas the other one can be gauged away in a complete analogy  with the higher superspin systems constructed in the present paper.
On the other hand, 
since in the present paper we deal with irreducible higher spin supermultiplets,
it would be also interesting to find a connection 
with the BRST formulation \cite{Buchbinder:2001bs}--\cite{Buchbinder:2007vq} for irreducible higher spin models as well.

\item It would be of particular interest to consider  cubic and possibly higher order   Lagrangians on AdS${}_3$ 
in the ``metric-like" approach following 
the lines of \cite{Buchbinder:2006eq}
(see also \cite{Boulanger:2011qt}--\cite{Sleight:2016dba} for cubic interactions of higher spin fields on AdS background).
One can investigate also a possibility of constructing cubic and higher order Lagrangians
for analogous systems with 3D
${\cal N}=2$ supersymmetry which in principle can be more restrictive on the level
of interactions comparing to  ${\cal N}=1$ supersymmetry considered here.

\item Vasiliev's higher spin gauge theory \cite{Vasiliev:2003ev,Vasiliev:1990en} 
was extended to superspace \cite{ESS,Engquist:2002gy}, 
although no analysis appeared as to  whether this approach reproduces 
the off--shell higher spin supermultiplets in AdS${}_4$ \cite{KS94} at the linearised level. 
Studying such issues in the 3D case seems to be less involved  than
in four dimensions.

\item  There has been much interest to higher spin (super)conformal 
field theories in three dimensions 
\cite{KO,K16,PopeTownsend,FL,Nilsson:2013tva,Nilsson:2015pua,Henneaux:2015cda,Linander:2016brv}. We hope our results will be useful for formulating 
interacting higher spin superconformal theories. 

\item Higher spin gauge fields possess interesting patterns of duality, 
both in the bosonic (see \cite{Hinterbichler:2016fgl,Henneaux:2016zlu} 
and references therein) and supersymmetric cases 
\cite{K16}. It would be interesting to continue studying 
the duality aspects of higher spin gauge fields.

\end{itemize}

\subsection*{\bf Acknowledgments}
\noindent
We are grateful to Joseph Novak for comments on the manuscript.
SMK thanks the Galileo Galilei Institute for Theoretical Physics for the hospitality and the INFN for partial support during the completion of this work. 
MT would like to thank the Department of Mathematics, the  University of Auckland and the Faculty of Education, Science, Technology and Mathematics, the University of Canberra for their kind hospitality during the various stages of the project.
This work is supported in part by the Australian Research Council,
project No. DP160103633.

\setcounter{equation}0
\appendix
\numberwithin{equation}{section}

\section{Some useful identities}\label{Appendix B}

Our 3D notation and conventions correspond to those introduced in 
\cite{KPT-MvU,KLT-M11}. 
In particular, the spinor indices are  raised and lowered using
the $\sSL(2,{\mathbb R})$ invariant tensors
\bea
\ve_{\a\b}=\left(\begin{array}{cc}0~&-1\\1~&0\end{array}\right)~,\qquad
\ve^{\a\b}=\left(\begin{array}{cc}0~&1\\-1~&0\end{array}\right)~,\qquad
\ve^{\a\g}\ve_{\g\b}=\d^\a_\b
\eea
using the standard rule:
\be
\theta^\ga = \ve^{\ga \gb} \theta_\gb, \quad 
\theta_\ga = \ve_{\ga \gb} \theta^\gb~, 
\ee

The spinor covariant derivative of $\cN=1$ Minkowski superspace is
\bea
D_\a =\frac{\pa}{\pa \q^\a} + {\rm i}  (\g^m)_{\a\b}\, \q^\b \pa_m
=  \frac{\pa}{\pa \q^\a} + {\rm i}  \q^\b \pa_{\b\a}~,
\eea
and obeys the anti-commutation relation
\be \label{ALG}
\{ D_\ga , D_\gb \} =2 \ri \pl_{\ga \gb}~.
\ee
As a result of \p{ALG} we have the identities
\begin{subequations}
\bea 
\label{DD}
&D_\ga D_\gb = \ri \pl_{\ga \gb} + \frac{1}{2}\epsilon_{\ga \gb}D^2~, 
\\
 \label{iden2}
&D^\b D_\ga D_\b =0~, 
\\ 
\label{iden3}
 &
 D^2 D_\ga= -D_\ga D^2 =
 2\ri \pl_{\ga \gb}D^\gb~,
\\
&
 D^2 D^2 = - 4 \Box~,
\label{iden4}
\eea
\end{subequations}
where $D^2 = D^\a D_\a$ and $\Box= \pl^a \pl_a = -\hf \pl^{\a \b}\pl_{\a \b} $.
An important corollary of \eqref{iden2} is
\be \label{DDDD}
[D_\ga D_\gb, D_\gc D_\delta ]=0~.
\ee

As compared with the supersymmetry in four dimensions, 
the spinor covariant derivative possesses unusual conjugation properties.
Specifically, given an arbitrary superfield $F$ 
and $\bar{F}:={(F)}^*$ its complex conjugate, the following relation holds
\bea
{(D_\a F)}^*=-(-1)^{\e(F)}D_\a\bar{F} ~,
\eea
where $\e(F)$ denotes the  Grassmann parity of $F$. 

The supersymmetry generator is 
\bea
Q_\a ={\rm i}\, \frac{\pa}{\pa \q^\a} + (\g^m)_{\a\b}\, \q^\b \pa_m
= {\rm i}\, \frac{\pa}{\pa \q^\a} +   \q^\b \pa_{\b\a}~.
\eea
It anti-commutes with the spinor covariant derivative 
\bea
\{Q_\a, D_\b\}=0~.
\eea


\allowdisplaybreaks

\section{Some useful identities for the operators}\label{Appendix C}

The operators introduced in the subsection  \ref{OSCILLATORS}
obey some useful relations
\begin{alignat}{2}
\{ \gamma, \gamma^+ \} & = 2 \ri K_1 - D^2~, &\quad  \gamma^+ \gamma &=  \ri K_ 1+\frac{1}{2} ND^2~, \\
P^+ K_1 &= -\ri \gamma^{+2} P +  \gamma^+ N \Box~, &\quad
K_1 P& = -\ri P^+ \gamma^2 -  N \gamma \Box~, \\
K_1 \gamma & = -\ri \gamma^+ \gamma^2 - NP~,  &\quad \gamma^+ K_1 &= -\ri \gamma^{+ 2} \gamma + P^+ N ~,\\
\label{aaaa}
[K_1, P]&=\Box \gamma~, &\quad [K_1, P^+]&= \gamma^+ \Box ~,\\
PD^2 & = -2 \ri \Box \gamma~, &\quad D^2 P &= 2 \ri \Box \gamma ~,\\
 \label{bbbb}
P^+D^2 &=-2 \ri \gamma^+ \Box~, &\quad D^2 P^+ &= 2 \ri \gamma^+ \Box~, \\
[\gamma, \gamma^{+2} ] &= - 2 \ri P^+~, &\quad [\gamma^2, \gamma^+] &= 2 \ri P ~,\\
\gamma  D^2& = -2 \ri P~, &\quad D^2 \gamma  &= 2 \ri P~, \\
 \gamma^+ D^2& = -2 \ri P^+~, &\quad
 D^2 \gamma^+ &= 2\ri P^+~, \\
[N, \gamma^+]&= \gamma^+~,  &\quad  [N, P^+]&= P^+~, \\
[N, \gamma] &= -\gamma~,  &\quad  [N, P]&= -P~, \\
[N,\Box]&= [N,D^2]= ~,0 &\quad [N, K_l]&=0~.
\end{alignat}
One has also the identity
\be \label{ll}
\gamma^{+2} \gamma^2 = -K_2 + N (N-1) \Box~,
\ee 
as well as ``reduction" rules for the operators $K_1$, $K_2$ and $K_l$ 
\bea \label{k1kl}
&&K_1 K_{l} = K_{l+1} +l  K_{l-1}(N-(l-1)) \Box ~,\\ \nonumber
&&K_2 K_l = K_{l+2} +2 l K_{l}(N-l) \Box + {l(l-1)} K_{l-2}(N-(l-1))(N-(l-2))\Box^2~,
\eea
where $K_0=1$.

\begin{footnotesize}

\end{footnotesize}

\end{document}